\documentclass[aps,showpacs,preprint,footinbib,preprintnumbers]{revtex4}

\def\ltap{\ \raise.3ex\hbox{$<$\kern-.75em\lower1ex\hbox{$\sim$}}\ }
\def\gtap{\ \raise.3ex\hbox{$>$\kern-.75em\lower1ex\hbox{$\sim$}}\ }
\def\gl{\ \raise.4ex\hbox{$>$\kern-.75em\lower1ex\hbox{$<$}}\ }

\usepackage[normalem]{ulem}  
\usepackage[dvips]{color} 
\usepackage{amsmath,amssymb}
\usepackage{booktabs}
\usepackage{mathrsfs}
\usepackage[dvipdfmx]{graphicx}
\usepackage{graphicx}
\usepackage{multirow}

\renewcommand\sout{\bgroup \color{red} \ULdepth=-.5ex \ULset}

\newcommand{\Ex}[2]{\ifmmode{#1\times10^{#2}}\else{$#1\times10^{#2}$}\fi}

\begin{document}

\title{Exotic mesons with hidden bottom near thresholds}
\author{S.~Ohkoda$^1$}
\author{Y.~Yamaguchi$^1$}
\author{S.~Yasui$^2$}
\author{K.~Sudoh$^3$}
\author{A.~Hosaka$^1$}
\affiliation{$^1$Research Center for Nuclear Physics (RCNP), 
Osaka University, Ibaraki, Osaka, 567-0047, Japan}
\affiliation{$^2$KEK Theory Center, Institute of Particle and Nuclear
Studies, High Energy Accelerator Research Organization, 1-1, Oho,
Ibaraki, 305-0801, Japan}
\affiliation{$^3$Nishogakusha University, 6-16, Sanbancho, Chiyoda,
Tokyo, 102-8336, Japan}

\begin{abstract}
We study  heavy hadron spectroscopy near open bottom
 thresholds. We employ B and
 $\mathrm{B}^{\ast}$ mesons as effective degrees of freedom near the
 thresholds, and consider meson exchange potentials between them. 
All possible composite states which can be constructed
 from the B and $\mathrm{B}^{\ast}$ mesons are studied up to the total
 angular momentum $J \le 2$. We consider, as exotic states, isosinglet states with
 exotic $J^{PC}$ quantum numbers and isotriplet states.  We solve
 numerically the Schr\"odinger equation with channel-couplings for each
 state. The masses of 
 twin resonances $\mathrm{Z}_{\mathrm b}(10610)$ and
 $\mathrm{Z}_{\mathrm b}(10650)$ recently found by
 Belle are reproduced.  We predict several
 possible bound and/or resonant states in other channels for future
 experiments.
\end{abstract}
\maketitle

\section{Introduction}

Exotic hadrons are studied extensively in recent hadron physics.
There have been many analyses which imply that they are multi-quark systems or hadronic molecules.
In strangeness sector, there are several candidates of exotic hadrons, such as $\mathrm{f}_{0}(980)$, $\mathrm{a}_{0}(980)$, $\Lambda(1405)$ and so on.
The scalar mesons $\mathrm{f}_{0}(980)$ and $\mathrm{a}_{0}(980)$ may be regarded as
tetra-quark systems or $\mathrm{K}\bar{\mathrm{K}}$ molecules \cite{Weinstein:1990gu,Oller:1998zr}. 
$\Lambda(1405)$ is considered to be generated dynamically by
$\bar{\mathrm{K}}\mathrm{N}$ and $\pi \Sigma$ \cite{Hyodo:2011ur}. 
In charm and bottom sectors, recently many candidates of exotic hadrons have been reported in experiments and also actively discussed in theoretical studies \cite{Brambilla:2004wf, Swanson:2006st,Voloshin:2007dx,Nielsen:2009uh,Brambilla:2010cs}.
$\mathrm{D}_{\mathrm{s}}(2317)$ and $\mathrm{D}_{\mathrm{s}}(2460)$ may be tetra-quarks or $\mathrm{KD}$ molecules.
$\mathrm{X}(3872)$, $\mathrm{Y}(4260)$, $\mathrm{Z}(4050)^{\pm}$,
$\mathrm{Z}(4250)^{\pm}$, $\mathrm{Z}(4430)^{\pm}$ and so on are also
candidates of exotics states. 
Especially $\mathrm{Z}(4050)^{\pm}$, $\mathrm{Z}(4250)^{\pm}$ and
$\mathrm{Z}(4430)^{\pm}$ cannot be simple charmonia
($\mathrm{c}\bar{\mathrm{c}}$) because they are electrically charged. 
There are also exotic hadrons in bottom flavors.
$\mathrm{Y}_{\mathrm{b}}$ is the first candidate of exotic bottom hadrons.
More recently, $\mathrm{Z}_{\mathrm{b}}(10610)^{\pm}$ and
$\mathrm{Z}_{\mathrm{b}}(10650)^{\pm}$ with isospin one have been
reported by Belle~\cite{Collaboration:2011gja,Belle:2011aa}.
The reported masses and widths of the two resonances are
$M(\mathrm{Z}_{\mathrm{b}}(10610)) = 10607.2 \pm 2.0$ MeV, 
$\Gamma (\mathrm{Z}_{\mathrm{b}}(10610)) = 18.4 \pm 2.4$ MeV and
$M (\mathrm{Z}_{\mathrm{b}}(10650)) = 10652.2 \pm 1.5$ MeV,
$\Gamma (\mathrm{Z}_{\mathrm{b}}(10650))=(11.5 \pm 2.2)$ MeV.
They also cannot be simple bottomonia ($\mathrm{b}\bar{\mathrm{b}}$) because they are electrically charged.

Well below the thresholds in the heavy quark systems,
quarkonia are  described by heavy quark degrees of freedom,
$\mathrm{Q}$ and  $\bar{\mathrm{Q}}$ ($\mathrm{Q}=\mathrm{b}$, $\mathrm{c}$).
Above the thresholds, however, it is a non-trivial problem whether the
resonant states are still explained by the quarkonium picture.
Clearly, a pair of heavy quark and anti-quark ($\mathrm{Q}\bar{\mathrm{Q}}$) are not sufficient
 effective degrees of freedom  to form the
resonances, because they are affected by the scattering states of the
two open heavy mesons. 
Indeed, many resonant states are found around the
thresholds in experiments. However they do not fit into the ordinary 
classification scheme of hadrons, such as the quark model calculation. 
Properties for masses, decay widths, branching ratios, and so forth, are
not predicted by the simple quarkonium picture \cite{Brambilla:2010cs}.
Therefore it is necessary to introduce components other than 
$\mathrm{Q}\mathrm{\bar{Q}}$ as effective
degrees of freedom around the thresholds. 

Instead of the dynamics of $\mathrm{Q}\mathrm{\bar{Q}}$, in the present
paper, we study the dynamics described by a pair of
a pseudoscalar meson $\mathrm{P} \sim
(\bar{\mathrm{Q}}\mathrm{q})_{\mathrm{spin\,0}}$ or a vector meson
$\mathrm{P}^{\ast} \sim (\bar{\mathrm{Q}}\mathrm{q})_{\mathrm{spin\,1}}$
($\mathrm{q}=\mathrm{u}$, $\mathrm{d}$) and their anti-mesons $\bar{\mathrm{P}}$ or
$\bar{\mathrm{P}}^{\ast}$, which are relevant hadronic degrees of
freedom around the thresholds.
In the following, we introduce the notation
$\mathrm{P}^{(\ast)}$ for $\mathrm{P}$
or $\mathrm{P}^{\ast}$ for simplicity.
 We discuss the possible
existence of the $\mathrm{P}^{(\ast)}\bar{\mathrm{P}}^{(\ast)}$ bound and/or resonant
states near the thresholds.
An interesting feature is that the pseudoscalar $\mathrm{P}$ meson and the vector $\mathrm{P}^{\ast}$ meson become degenerate in mass in the heavy quark limit ($M_{\mathrm{Q}} \rightarrow \infty$).
The mass degeneracy originates from the suppression of the Pauli term
in the magnetic gluon sector in QCD, which is 
the quantity of order
${\cal O}(1/M_{\mathrm{Q}})$ with heavy quark mass
$M_{\mathrm{Q}}$ \cite{Isgur:1991wq,Manohar:2000dt}.
Therefore, the effective degrees of freedom at the threshold are given,
not only by $\mathrm{P}\bar{\mathrm{P}}$, but also by
combinations, such as
$\mathrm{P}^{\ast}\bar{\mathrm{P}}$, $\mathrm{P}\bar{\mathrm{P}}^{\ast}$
and $\mathrm{P}^{\ast}\bar{\mathrm{P}}^{\ast}$. 
Because $\mathrm{P}^{(\ast)}$ 
includes a heavy anti-quark $\bar{\mathrm{Q}}$ and a light quark
$\mathrm{q}$,  
the Lagrangian of $\mathrm{P}$ and $\mathrm{P}^{\ast}$ meson systems is
given with respecting the heavy quark symmetry (spin symmetry) and
chiral symmetry
\cite{Manohar:2000dt,Burdman:1992gh,Wise:1992hn,Yan:1992gz,Nowak:1992um,Bardeen:1993ae,Badalian:2007yr,Casalbuoni:1996pg,Bardeen:2003kt,Matsuki:2007zza}. 

A new degree of freedom which does not exist in the
$\mathrm{Q}\bar{\mathrm{Q}}$ systems but does only in the
$\mathrm{P}^{(\ast)}\bar{\mathrm{P}}^{(\ast)}$ systems is an
isospin. 
Then, there appears one pion exchange potential (OPEP) between
$\mathrm{P}^{(\ast)}$ and $\bar{\mathrm{P}}^{(\ast)}$ mesons at long
distances of order $1/m_{\pi}$ with pion mass $m_{\pi}$.
What is interesting in the OPEP between $\mathrm{P}^{(\ast)}$ and
$\mathrm{\bar{P}}^{(\ast)}$ is that it causes a mixing between
states of
different angular momentum, such as $L$ and $L \pm 2$,
 through its tensor component.
Therefore, it  is expected that the
$\mathrm{P}^{(\ast)}\bar{\mathrm{P}}^{(\ast)}$ systems behave 
 differently from the quarkonium systems. 
In reality in addition to the one pion exchange dominated at long distances, there are multiple pion ($\pi\pi$, $\pi \pi \pi$, etc.) exchange, heavy meson ($\rho$, $\omega$, $\sigma$, etc.) exchange at short distances as well.
With these potentials, we solve the two-body Schr\"odinger equation
with channel-couplings and  discuss the
existence of bound and/or resonant states of
$\mathrm{P}^{(\ast)}\bar{\mathrm{P}}^{(\ast)}$.

In this paper we study  $\mathrm{P}^{(\ast)}\bar{\mathrm{P}}^{(\ast)}$
systems, with exotic quantum numbers
which cannot be accessed by  quarkonia. 
The first group is for isosinglet states with  $I=0$. 
We recall  that the possible $J^{PC}$ of
quarkonia are $J^{PC}=0^{-+}$ ($\eta_{\mathrm{b}}$), $0^{++}$
($\chi_{\mathrm{b}0}$) for $J=0$, $J^{--}$ ($\Upsilon$), $J^{+-}$
($\mathrm{h}_{\mathrm{b}}$), $J^{++}$ ($\chi_{\mathrm{b}1}$) for odd $J
\ge 1$, and $J^{--}$, $J^{-+}$, $J^{++}$ ($\chi_{\mathrm{b}2}$) for even
$J \ge 2$, where examples of bottomonia are shown in the parentheses.
However, there cannot be $J^{PC}=0^{--}$ and $0^{+-}$, $J^{-+}$ with odd $J \ge 1$, and $J^{+-}$ with even $J \ge 2$ in the quarkonia.
These quantum numbers are called exotic $J^{PC}$, and it has been
discussed that they are the signals for exotics including
 the $\mathrm{P}^{(\ast)}\bar{\mathrm{P}}^{(\ast)}$ systems and glueballs.
The second group  is for  isospin triplet states with  $I=1$.
It is obvious that the quarkonia themselves cannot be isotriplet.
To have a finite isospin, there must be additional light quark degrees of freedom \cite{Dubynskiy:2008mq}.
In this regard, $\mathrm{P}^{(\ast)}$ and $\bar{\mathrm{P}}^{(\ast)}$
mesons have isospin half, and therefore the
$\mathrm{P}^{(\ast)}\bar{\mathrm{P}}^{(\ast)}$ composite systems
can be  isospin triplet. 
We observe that, near the thresholds, the
$\mathrm{P}^{(\ast)}\bar{\mathrm{P}}^{(\ast)}$ systems can access to
more variety of quantum numbers than  the
$\mathrm{Q}\bar{\mathrm{Q}}$ systems. 
In this paper, we focus on the bottom sector ($\mathrm{P}=\mathrm{B}$ and $\mathrm{P}^{\ast}=\mathrm{B}^{\ast}$), because the heavy quark symmetry works better than the charm sector.

In the previous works, Ericson and Karl estimated the 
OPEP in hadronic molecules within strangeness sector and indicated
the importance of tensor interaction in this system \cite{Ericson:1993wy}.
 T\"ornqvist analyzed one pion exchange force between  two
mesons for many possible quantum numbers in \cite{Tornqvist:1991ks,Tornqvist:1993ng}.
Inspired by the discovery of $\mathrm{X}(3872)$, the hadronic molecular
model has  
been developed by many authors \cite{Swanson:2006st,Nielsen:2009uh,Lee:2009hy,Thomas:2008ja,Liu:2008tn,Liu:2008fh,Suzuki:2005ha}.
For $\mathrm{Z}_{\mathrm{b}}$'s many works have already been done since 
the Belle's discovery. 
As candidates of exotic states, molecular structure has been
studied~\cite{Zhang:2011jja,Bugg:2011jr,Yang:2011rp,Nieves:2011zz,Sun:2011uh,Cleven:2011gp,Mehen:2011yh},
and also tetraquark structure~\cite{Guo:2011gu,Cui:2011fj,Navarra:2011xa,Ali:2011vy,Karliner:2011yb,Ali:2011ug}. 
The existence of $\mathrm{Z}_{\mathrm{b}}$'s has also been 
investigated in the decays of
$\Upsilon(5S)$~\cite{Chen:2011zv,Chen:2011pv,Chen:2011xk,Chen:2011pu}.
Our study based on the molecular picture of 
$\mathrm{P}^{(\ast)}\bar{\mathrm{P}}^{(\ast)}$
differs from the previous works in that we completely take into
account the degeneracy of pseudoscalar meson $\mathrm{B}$ and a vector
meson $\mathrm{B}^{\ast}$ due to the heavy quark symmetry,
and fully consider channel-couplings of $\mathrm{B}^{(\ast)}$ and
$\bar{\mathrm{B}}^{(\ast)}$.
 In the previous publications, the low lying molecular states around
 $\mathrm{Z}_{\mathrm{b}}$'s which can be produced from the decay of
 $\Upsilon(5S)$ 
were studied systematically and qualitatively~\cite{Bondar:2011ev,Voloshin:2011qa}.
Our present work covers them also.

This paper is organized as followings.
In section 2, we introduce (i) the $\pi$ exchange potential and (ii) the $\pi \rho\, \omega$ potential between $\mathrm{B}^{(\ast)}$ and $\bar{\mathrm{B}}^{(\ast)}$ mesons.
To obtain the potentials, we respect the heavy quark symmetry for
the $\mathrm{B}^{(\ast)}\mathrm{B}^{(\ast)}\pi$,
$\mathrm{B}^{(\ast)}\mathrm{B}^{(\ast)}\rho$ and
$\mathrm{B}^{(\ast)}\mathrm{B}^{(\ast)}\omega$ vertices. 
In section 3, we classify all the possible states composed by a pair of
$\mathrm{B}^{(\ast)}$ and $\bar{\mathrm{B}}^{(\ast)}$ mesons with
exotic quantum numbers $I^{G}(J^{PC})$
with isospin $I$, $G$-parity, total angular momentum $J$, parity $P$
and charge conjugation $C$. ($C$ in $I=1$ is defined only for states of $I_{z}=0$.)
In section 4, we solve numerically the Schr\"odinger equations with channel-couplings and discuss the bound and/or resonant states of the $\mathrm{B}^{(\ast)}\bar{\mathrm{B}}^{(\ast)}$ systems.
We employ the hadronic molecular picture and only consider the
$\mathrm{B}^{(\ast)}\bar{\mathrm{B}}^{(\ast)}$ states.
In practice, there are bottomonium and light meson states which couple to these states.
The effect of these couplings as quantum corrections is estimated in section 5. 
In section 6, we discuss the possible decay modes of these states.
Section 7 is devoted to summary.

\section{Interactions with heavy quark symmetry}

 $\mathrm{B}^{(\ast)}$ mesons have a heavy anti-quark $\bar{\mathrm{b}}$ and a light quark $\mathrm{q}=\mathrm{u}$, $\mathrm{d}$.
The dynamics of the $\mathrm{B}^{(\ast)}\bar{\mathrm{B}}^{(\ast)}$
systems is given by the two symmetries: the heavy quark symmetry
for heavy quarks and chiral symmetry for light quarks. 
These two  symmetries 
 provide the vertices of $\pi$ meson and of vector
 meson ($v=\rho$, $\omega$) with open heavy flavor (bottom) mesons
 $P$  and $P^{*}$ 
($P$ for $\mathrm{B}$ and $P^{*}$ for $\mathrm{B^*}$)
\begin{eqnarray}
{\cal L}_{\pi H H} &=& g \, \mbox{tr} \bar{H}_{a}H_{b}\gamma_{\nu}\gamma_{5}A_{ba}^{\nu},
\label{eq:piHH} \\
{\cal L}_{v H H} &=& -i\beta \mbox{tr} \bar{H}_{a} H_{b} v^{\mu}
(\rho_{\mu})_{ba}  
  + i\lambda \mbox{tr}\bar{H}_{a} H_{b}\sigma_{\mu \nu}
F_{\mu \nu}(\rho)_{ba}\, , 
\label{eq:vHH}
\end{eqnarray}
where the multiplet field $H$ containing $P$  and $P^{*}$  is defined by
\begin{eqnarray}
H_{a} = \frac{1+/\hspace{-0.5em}v}{2} \left[ P_{a\mu}^{*} \gamma^{\mu} - P_{a} \gamma_{5} \right],
\end{eqnarray}
with the four velocity $v_{\mu}$ of the heavy mesons \cite{Isgur:1991wq}.
The conjugate field is defined by $\bar{H}_{a} = \gamma_{0} H_{a}^{\dag} \gamma_{0}$, and the index $a$ denotes up and down flavors.
The axial current is given by
$
A_{\mu} \simeq \frac{i}{f_{\pi}} \partial_{\mu}\hat{\pi}
$
with 
\begin{eqnarray}
\hat{\pi} = \left(
\begin{array}{cc}
 \frac{\pi^{0}}{\sqrt{2}} & \pi^{+} \\
 \pi^{-} & -\frac{\pi^{0}}{\sqrt{2}}
\end{array}
\right),
\end{eqnarray}
where $f_{\pi}=135$ MeV is the pion decay constant.
The coupling constant $|g|=0.59$ for $\pi P P^{*}$ is determined with reference to
the observed decay width  $\Gamma = 96$ keV for $\mathrm{D}^{*}
\rightarrow \mathrm{D}\pi$ \cite{Amsler:2008zzb}, assuming that the charm quark is sufficiently heavy. 
The coupling constant $g$ for $\pi \mathrm{B} \mathrm{B}^{*}$ would be
different from the one for $\pi \mathrm{D} \mathrm{D}^{*}$ because of $1/m_{\mathrm{Q}}$ corrections with the heavy quark mass $m_{\mathrm{Q}}$ \cite{Cheng:1993gc}.
However the lattice simulation in the heavy quark limit suggests a
similar value  as adopted above \cite{Ohki:2008py},
 allowing us to use the common value for $\mathrm{D}$ and $\mathrm{B}$.
The coupling of $\pi P^{*}P^{*}$, which is difficult to access from 
experiments, is also fixed thanks to the heavy quark symmetry.
Note that the coupling of $\pi P P$ does not exist due to the parity conservation.
The coupling constants $\beta$ and $\lambda$ are determined by the 
radiative decays of $\mathrm{D^{\ast}}$ meson and semileptonic decays of
$\mathrm{B}$ meson with vector meson dominance as $\beta =0.9$ and $\lambda= 0.56$ $\mathrm{GeV}^{-1}$ by following Ref.~\cite{Isola:2003fh}.
The vector ($\rho$ and $\omega$) meson field 
 is defined by
\begin{eqnarray}
 & \rho_{\mu} = i\displaystyle{\frac{g_V}{\sqrt{2}}\hat{\rho_{\mu}}} \, , 
\end{eqnarray}
with
\begin{eqnarray}
 & \hat{\rho_{\mu}} =  \displaystyle{\left(
\begin{array}{cc}
 \frac{\rho^{0}}{\sqrt{2}} + \frac{\omega}{\sqrt{2}}& \rho^{+} \\
 \rho^{-} & -\frac{\rho^{0}}{\sqrt{2}} + \frac{\omega}{\sqrt{2} }
\end{array}
\right)}_{\mu} \, , 
\end{eqnarray}
and its field tensor by
\begin{eqnarray}
F_{\mu \nu}(\rho) = \partial_{\mu}\rho_{\nu} -\partial_{\nu}\rho_{\mu}
 +[\rho_{\mu},\rho_{\nu}] \, ,
\end{eqnarray}
where $g_V = 5.8$ is the coupling constant for $\rho \rightarrow \pi
\pi$ decay.

From Eq.~(\ref{eq:piHH}), we obtain the $\pi PP^*$ and $\pi P^* P^*$
vertices
\begin{eqnarray}
 {\cal L}_{\pi PP^*} &=& 
 2 \frac{g}{f_\pi}(P^\dagger_a P^\ast_{b\,\mu}+P^{\ast\,\dagger}_{a\,\mu}P_b)\partial^\mu\hat{\pi}_{ab}
 \, , 
 \label{eq:piPP*}\\
 {\cal L}_{\pi P^*P^*} &=& 
 2 i \frac{g}{f_\pi}\epsilon^{\alpha \beta \mu \nu} v_{\alpha}
 P^{\ast\,\dagger}_{a\,\beta}P^\ast_{b\,\mu}\partial_{\nu}
 \hat{\pi}_{ab}
 \, .
\label{eq:piP*P*}
\end{eqnarray}
The $\pi \bar{P} \bar{P}^{\ast}$ and $\pi \bar{P}^{\ast} \bar{P}^{\ast}$
vertices are obtained by changing the sign of the $\pi P P^{\ast}$ and
$\pi P^{\ast} P^{\ast}$ vertices in Eqs.~(\ref{eq:piPP*}) and
(\ref{eq:piP*P*}). 
%
Similarly, from Eq.~(\ref{eq:vHH}) we derive the $vPP$, $vPP^{\ast}$ and $vP^{\ast}P^{\ast}$ vertices ($v=\rho$, $\omega$) as
\begin{eqnarray}
 {\cal L}_{vPP} &=&
-\sqrt{2}\beta g_V P_b P^{\dagger}_a v \cdot \hat{\rho}_{ba}
 \, , 
\label{eq:vPP} \\
{\cal L}_{vPP^*} &=& 
-2\sqrt{2}\lambda g_V v_{\mu}\epsilon^{\mu \nu \alpha \beta}
\left(P^\dagger_a
P^\ast_{b\,\beta}-P^{\ast\,\dagger}_{a\,\beta}P_b\right)
\partial_\nu(\hat{\rho}_\alpha)_{ba}
\, ,
\label{eq:vPP*} \\
{\cal L}_{vP^*P^*} &=&
\sqrt{2} \beta g_V P^*_b P^{*\dagger}_a v \cdot \hat{\rho}_{ba}\nonumber \\
&&+i2 \sqrt{2}\lambda g_V
P^{\ast\,\dagger}_{a\,\mu}P^\ast_{b\,\nu}
(\partial^\mu(\hat{\rho}^\nu)_{ba}-\partial^\nu(\hat{\rho}^\mu)_{ba}) \, .
\label{eq:vP*P*}
\end{eqnarray}
Due to the $G$-parity, the signs of vertices for $v\bar{P}\bar{P}$, $v\bar{P}\bar{P}^{\ast}$ and
$v\bar{P}^{\ast}\bar{P}^{\ast}$ are opposite to those of $vPP$, $vPP^*$
and $vP^* P^*$, respectively, for $v=\omega$, while they are the same  
for $v=\rho$. 

It is important that the scatterings
$P^{(\ast)}\bar{P}^{(\ast)} \rightarrow
P^{(\ast)}\bar{P}^{(\ast)}$ include not only diagonal components
$P\bar{P}^{\ast} \rightarrow
P^{\ast}\bar{P}$ and
$P^{\ast}\bar{P}^{\ast} \rightarrow
P^{\ast}\bar{P}^{\ast}$ but also off-diagonal
components $P\bar{P} \rightarrow
P^{\ast}\bar{P}^{\ast}$ and
$P\bar{P}^{\ast} \rightarrow
P^{\ast}\bar{P}^{\ast}$. 
The OPEPs for 
$P\bar{P}^{\ast} \rightarrow
P^{\ast}\bar{P}$ and
$P^{\ast}\bar{P}^{\ast} \rightarrow
P^{\ast}\bar{P}^{\ast}$ are given from the vertices (\ref{eq:piPP*}) and (\ref{eq:piP*P*}) in the heavy quark
limit as 
\begin{eqnarray}
V^{\pi}_{P_{1}\bar{P}_{2}^{\ast} \rightarrow P_{1}^{\ast}\bar{P}_{2}} &\!=\!&
 -\left( \sqrt{2} \frac{g}{f_{\pi}} \right)^{2} \frac{1}{3} \left[ \vec{\varepsilon}_{1}^{\,\ast} \!\cdot\! \vec{\varepsilon}_{2} \, C(r;m_{\pi}) \!+\! S_{\varepsilon_{1}^{\ast},\varepsilon_{2}} \, T(r;m_{\pi}) \right]  \vec{\tau}_{1}  \!\cdot\! \vec{\tau}_{2}, \label{eq:pot_BB*B*B}  \\
V^{\pi}_{P_{1}^{\ast}\bar{P}_{2}^{\ast} \rightarrow P_{1}^{\ast}\bar{P}_{2}^{\ast}} &\!=\!&
 -\left( \sqrt{2} \frac{g}{f_{\pi}} \right)^{2} \frac{1}{3} \left[ \vec{T}_{1} \!\cdot\! \vec{T}_{2} \, C(r;m_{\pi}) \!+\! S_{T_{1},T_{2}} \, T(r;m_{\pi}) \right]  \vec{\tau}_{1}  \!\cdot\! \vec{\tau}_{2}, \label{eq:pot_B*B*B*B*}
\end{eqnarray}
and the OPEPs for  $P\bar{P} \rightarrow P^{\ast}\bar{P}^{\ast}$ and $P\bar{P}^{\ast} \rightarrow P^{\ast}\bar{P}^{\ast}$ are given as
\begin{eqnarray}
V^{\pi}_{P_{1}\bar{P}_{2} \rightarrow P_{1}^{\ast}\bar{P}_{2}^{\ast}} &\!=\!&
 -\left( \sqrt{2} \frac{g}{f_{\pi}} \right)^{2} \frac{1}{3} \left[ \vec{\varepsilon}_{1}^{\,\ast} \!\cdot\! \vec{\varepsilon}_{2}^{\, \ast} \, C(r;m_{\pi}) \!+\! S_{\varepsilon_{1}^{\ast},\varepsilon_{2}^{\ast}} \, T(r;m_{\pi}) \right] \vec{\tau}_{1}  \!\cdot\! \vec{\tau}_{2}, \label{eq:pot_BBB*B*} \\
V^{\pi}_{P_{1}\bar{P}_{2}^{\ast} \rightarrow P_{1}^{\ast}\bar{P}_{2}^{\ast}} &\!=\!&
 \left( \sqrt{2} \frac{g}{f_{\pi}} \right)^{2} \frac{1}{3} \left[ \vec{\varepsilon}_{1}^{\,\ast} \!\cdot\! \vec{T}_{2} \, C(r;m_{\pi}) \!+\! S_{\varepsilon_{1}^{\ast},T_{2}} \, T(r;m_{\pi}) \right]  \vec{\tau}_{1}  \!\cdot\! \vec{\tau}_{2}. \label{eq:pot_BB*B*B*}
\end{eqnarray}
Here three polarizations are possible for $P^{*}$ as defined by
$\vec{\varepsilon}^{\hspace{0.2em}(\pm)} \!=\! \left(\mp 1/\sqrt{2}, \pm i/\sqrt{2}, 0 \right)$ and
$\vec{\varepsilon}^{\hspace{0.2em}(0)} \!=\! \left(0, 0, 1\right)$,
and the spin-one operator $\vec{T}$ is defined by $T_{\lambda' \lambda}^{i}=i \varepsilon^{ijk} \varepsilon_{j}^{(\lambda')\dag} \varepsilon_{k}^{(\lambda)}$.
As a convention, we assign $\vec{\varepsilon}^{\,(\lambda)}$ for an
incoming vector particle and $\vec{\varepsilon}^{\,(\lambda)\ast}$ for
an outgoing vector particle.
Here $\vec{\tau}_{1}$ and $\vec{\tau}_{2}$ are isospin operators for $P^{(\ast)}_{1}$ and $\bar{P}^{(\ast)}_{2}$. We define the tensor operators 
\begin{eqnarray}
S_{\varepsilon_{1}^{\ast},\varepsilon_{2}} &=& 3 ( \vec{\varepsilon}^{\,(\lambda_{1})\ast} \!\cdot\!\hat{r} ) ( \vec{\varepsilon}^{\,(\lambda_{2})} \!\cdot\!\hat{r} ) -  \vec{\varepsilon}^{\,(\lambda_{1})\ast} \!\cdot\! \vec{\varepsilon}^{\,(\lambda_{2})}, \\
S_{T_{1},T_{2}} &=& 3 ( \vec{T}_{1} \!\cdot\!\hat{r} ) ( \vec{T}_{2} \!\cdot\!\hat{r} ) - \vec{T}_{1} \!\cdot\! \vec{T}_{2}, \\
S_{\varepsilon_{1}^{\ast},\varepsilon_{2}^{\ast}} &=& 3 ( \vec{\varepsilon}^{\,(\lambda_{1})\ast} \!\cdot\!\hat{r} ) ( \vec{\varepsilon}^{\,(\lambda_{2})\ast} \!\cdot\!\hat{r} ) -  \vec{\varepsilon}^{\,(\lambda_{1})\ast} \!\cdot\! \vec{\varepsilon}^{\,(\lambda_{2})\ast}, \\
S_{\varepsilon_{1}^{\ast},T_{2}} &=& 3 ( \vec{\varepsilon}^{\,(\lambda_{1})\ast} \!\cdot\!\hat{r} ) ( \vec{T}_{2} \!\cdot\!\hat{r} ) -  \vec{\varepsilon}^{\,(\lambda_{1})\ast} \!\cdot\! \vec{T}_{2}.
\end{eqnarray} 
The $\rho$  meson exchange potentials are derived by
using the same notation of the OPEPs and the vertices in Eqs.~(\ref{eq:vPP})-(\ref{eq:vP*P*}),
\begin{eqnarray}
V^{v}_{P_{1}\bar{P}_{2} \rightarrow P_{1}\bar{P}_{2}} &\!=\!&
 \left( \frac{\beta g_V}{2m_{v}} \right)^{2} \frac{1}{3}  C(r;m_{v})
 \vec{\tau}_{1}  \!\cdot\! \vec{\tau}_{2}, \label{eq:rhopot_BBBB}  \\
V^{v}_{P_{1}\bar{P}_{2}^{\ast} \rightarrow 
P_{1}\bar{P}_{2}^{\ast}} &\!=\!&
 \left( \frac{\beta g_V}{2m_{v}} \right)^{2} \frac{1}{3}  C(r;m_{v})  \vec{\tau}_{1}  \!\cdot\! \vec{\tau}_{2}, \label{eq:rhopot_BB*BB*}  \\
 V^{v}_{P_{1}\bar{P}_{2}^{\ast} \rightarrow
  P_{1}^{\ast}\bar{P}_{2}} &\!=\!& 
 \left( 2\lambda g_V \right)^{2} \frac{1}{3} \left[
  2\vec{\varepsilon}_{1}^{\,\ast} \!\cdot\! \vec{\varepsilon}_{2} \,
  C(r;m_{v}) \!-\! S_{\varepsilon_{1}^{\ast},\varepsilon_{2}} \, T(r;m_{v})
					     \right]  \vec{\tau}_{1}
 \!\cdot\! \vec{\tau}_{2}, \label{eq:rhopot_BB*B*B}  \\
V^{v}_{P_{1}^{\ast}\bar{P}_{2}^{\ast} \rightarrow P_{1}^{\ast}\bar{P}_{2}^{\ast}} &\!=\!&
 \left(  2\lambda g_V \right)^{2} \frac{1}{3} \left[ 2\vec{T}_{1}
  \!\cdot\! \vec{T}_{2} \, C(r;m_{v}) \!-\! S_{T_{1},T_{2}} \,
  T(r;m_{v}) \right]  \vec{\tau}_{1}  \!\cdot\! \vec{\tau}_{2} \nonumber \\ 
&& + \left( \frac{\beta g_V}{2m_{v}} \right)^{2} \frac{1}{3}  C(r;m_{v})
 \vec{\tau}_{1}  \!\cdot\! \vec{\tau}_{2},  \label{eq:rhopot_B*B*B*B*} \\
 V^{v}_{P_{1}\bar{P}_{2} \rightarrow P_{1}^{\ast}\bar{P}_{2}^{\ast}} &\!=\!&
 \left(  2\lambda g_V  \right)^{2} \frac{1}{3} \left[2
  \vec{\varepsilon}_{1}^{\,\ast} \!\cdot\! \vec{\varepsilon}_{2}^{\,
  \ast} \, C(r;m_{v}) \!-\!
  S_{\varepsilon_{1}^{\ast},\varepsilon_{2}^{\ast}} \, T(r;m_{v})
					       \right] \vec{\tau}_{1}
 \!\cdot\! \vec{\tau}_{2}, \label{eq:rhopot_BBB*B*} \\
V^{v}_{P_{1}\bar{P}_{2}^{\ast} \rightarrow P_{1}^{\ast}\bar{P}_{2}^{\ast}} &\!=\!&
 -\left(  2\lambda g_V \right)^{2} \frac{1}{3} \left[ 2\vec{\varepsilon}_{1}^{\,\ast} \!\cdot\! \vec{T}_{2} \, C(r;m_{v}) \!-\! S_{\varepsilon_{1}^{\ast},T_{2}} \, T(r;m_{v}) \right]  \vec{\tau}_{1}  \!\cdot\! \vec{\tau}_{2}, \label{eq:rhopot_BB*B*B*}
\end{eqnarray}
for $v=\rho$. The $\omega$ exchange potentials are obtained by changing the overall sign from the above equations with $v=\omega$ and by removing the isospin factor $\vec{\tau}_{1}  \!\cdot\! \vec{\tau}_{2}$.

To estimate the size effect of mesons, we introduce a form
factor $(\Lambda^{2}-m_{h}^{2})/(\Lambda^{2}+\vec{q}^{\,\,2})$ in the
momentum space at vertices of 
$h PP$, $h PP^{\ast}$ and $h P^{\ast}P^{\ast}$ ($h=\pi$, $\rho$ and
$\omega$). 
Here $\vec{q}$ and $m_{h}$ are momentum and mass of the exchanged meson, and
$\Lambda$ is the cut-off parameter.
Then, $C(r;m_{h})$ and $T(r;m_{h})$  are defined as
\begin{eqnarray}
\hspace{-3em}&&C(r;m_{h}) \!=\! \int \frac{\mbox{d}^{3}\vec{q}}{(2\pi)^3} \frac{m_{h}^{2}}{\vec{q}^{\,\,2}+m_{h}^{2}} 
 e^{i\vec{q} \cdot \vec{r}} \, 
F(\vec{q};m_{h}), \\
\hspace{-3em}&&T(r;m_{h}) S_{12}(\hat{r}) \!=\! \int \frac{\mbox{d}^{3}\vec{q}}{(2\pi)^3} \frac{- \vec{q}^{\,\,2}}{\vec{q}^{\,\,2}+m_{h}^{2}} 
S_{12}(\hat{q})e^{i\vec{q} \cdot \vec{r}} F(\vec{q};m_{h}),
\end{eqnarray}
with $S_{12}(\hat{x}) \!=\! 3 (\vec{\sigma}_{1} \!\cdot\! \hat{x})
(\vec{\sigma}_{2} \!\cdot\! \hat{x}) - \vec{\sigma}_{1} \!\cdot\!
\vec{\sigma}_{2}$, and $F(\vec{q};m_{h})\!=\! (\Lambda^{2} \!-\!
m_{h}^{2})^{2}/(\Lambda^{2} \!+\! \vec{q}^{\,\,2})^{2}$.
The cut-off $\Lambda$ is determined from the size of $\mathrm{B}^{(\ast)}$ based on the quark model as discussed in Refs.~\cite{Yasui:2009bz,Yamaguchi:2011xb}.
There, the cut-off parameter is $\Lambda=1070$ MeV when the $\pi$ exchange potential is
employed, while $\Lambda=1091$ MeV when the $\pi \rho\, \omega$ potential is employed. 

As a brief summary, we emphasize again that, according to the heavy
quark symmetry, not only the $\mathrm{B}\bar{\mathrm{B}}^{\ast} \rightarrow
\mathrm{B}^{\ast}\bar{\mathrm{B}}$ and
$\mathrm{B}^{\ast}\bar{\mathrm{B}}^{\ast} \rightarrow
\mathrm{B}^{\ast}\bar{\mathrm{B}}^{\ast}$ transitions  but also the
$\mathrm{B}\bar{\mathrm{B}} \rightarrow
\mathrm{B}^{\ast}\bar{\mathrm{B}}^{\ast}$ and
$\mathrm{B}\bar{\mathrm{B}}^{\ast} \rightarrow
\mathrm{B}^{\ast}\bar{\mathrm{B}}^{\ast}$ transitions
become important as channel-couplings.
In the next section, we will see that the latter two transitions supply
the strong tensor force, through the channel mixing $\mathrm{B}$ and $\mathrm{B}^{\ast}$ as well as different angular momentum, such as $L$ and $L \pm 2$.

\section{Classification of the $\mathrm{B}^{(\ast)}\bar{\mathrm{B}}^{(\ast)}$ states}

\begin{table}[htdp]
\caption{ Various components of the
 $\mathrm{B}^{(\ast)}\bar{\mathrm{B}}^{(\ast)}$ states for several
 $J^{PC}$ ($J \le 2$). The exotic quantum numbers which cannot be
 assigned to bottomonia $\mathrm{b}\bar{\mathrm{b}}$ are indicated by
 $\surd$. The $0^{+-}$ state cannot be neither bottomonium nor
 $\mathrm{B}^{(\ast)}\bar{\mathrm{B}}^{(\ast)}$ states.}
\begin{center}
{\renewcommand\arraystretch{1.5}
\begin{tabular}{|c|c|c|c|}
\hline
$J^{PC}$ & components & \multicolumn{2}{c|}{exoticness} \\ 
\cline{3-4}
                         &                     & $I=0$ & $I=1$ \\ 
\hline
\hline
$0^{+-}$ & ------ & $\surd$ & $\surd$ \\ 
\hline
$0^{++}$ & $\mathrm{B}\bar{\mathrm{B}}(^{1}S_{0})$, $\mathrm{B}^{\ast}\bar{\mathrm{B}}^{\ast}(^{1}S_{0})$, $\mathrm{B}^{\ast}\bar{\mathrm{B}}^{\ast}(^{5}D_{0})$ & $\chi_{\mathrm{b}0}$ & $\surd$ \\ 
\hline
$0^{--}$ & $\frac{1}{\sqrt{2}} \left( \mathrm{B}\bar{\mathrm{B}}^{\ast}+\mathrm{B}^{\ast}\bar{\mathrm{B}} \right)(^{3}P_{0})$ & $\surd$ & $\surd$  \\ 
\hline
$0^{-+}$ & $\frac{1}{\sqrt{2}} \left( \mathrm{B}\bar{\mathrm{B}}^{\ast}-\mathrm{B}^{\ast}\bar{\mathrm{B}} \right)(^{3}P_{0})$, $\mathrm{B}^{\ast}\bar{\mathrm{B}}^{\ast}(^{3}P_{0})$ & $\eta_{\mathrm b}$ & $\surd$ \\ 
\hline
$1^{+-}$ & $\frac{1}{\sqrt{2}} \left( \mathrm{B}\bar{\mathrm{B}}^{\ast}-\mathrm{B}^{\ast}\bar{\mathrm{B}} \right) (^{3}S_{1})$, $\frac{1}{\sqrt{2}} \left( \mathrm{B}\bar{\mathrm{B}}^{\ast}-\mathrm{B}^{\ast}\bar{\mathrm{B}} \right) (^{3}D_{1})$, $\mathrm{B}^{\ast}\bar{\mathrm{B}}^{\ast}(^{3}S_{1})$, $\mathrm{B}^{\ast}\bar{\mathrm{B}}^{\ast}(^{3}D_{1})$ & $\mathrm{h}_{\mathrm b}$ & $\surd$ \\ 
\hline
$1^{++}$ & $\frac{1}{\sqrt{2}} \left( \mathrm{B}\bar{\mathrm{B}}^{\ast}+\mathrm{B}^{\ast}\bar{\mathrm{B}} \right) (^{3}S_{1})$, $\frac{1}{\sqrt{2}} \left( \mathrm{B}\bar{\mathrm{B}}^{\ast}+\mathrm{B}^{\ast}\bar{\mathrm{B}} \right)(^{3}D_{1})$, $\mathrm{B}^{\ast}\bar{\mathrm{B}}^{\ast}(^{5}D_{1})$ & $\chi_{\mathrm{b}1}$ & $\surd$ \\ 
\hline
$1^{--}$ & $\mathrm{B}\bar{\mathrm{B}}(^{1}P_{1})$, $\frac{1}{\sqrt{2}} \left( \mathrm{B}\bar{\mathrm{B}}^{\ast}+\mathrm{B}^{\ast}\bar{\mathrm{B}} \right)(^{3}P_{1})$, $\mathrm{B}^{\ast}\bar{\mathrm{B}}^{\ast}(^{1}P_{1})$, $\mathrm{B}^{\ast}\bar{\mathrm{B}}^{\ast}(^{5}P_{1})$, $\mathrm{B}^{\ast}\bar{\mathrm{B}}^{\ast}(^{5}F_{1})$ & $\Upsilon$ & $\surd$ \\ 
\hline
$1^{-+}$ & $\frac{1}{\sqrt{2}} \left( \mathrm{B}\bar{\mathrm{B}}^{\ast}-\mathrm{B}^{\ast}\bar{\mathrm{B}} \right)(^{3}P_{1})$, $\mathrm{B}^{\ast}\bar{\mathrm{B}}^{\ast}(^{3}P_{1})$ & $\surd$ & $\surd$ \\ 
\hline
$2^{+-}$ & $\frac{1}{\sqrt{2}} \left( \mathrm{B}\bar{\mathrm{B}}^{\ast}-\mathrm{B}^{\ast}\bar{\mathrm{B}} \right)(^{3}D_{2})$, $\mathrm{B}^{\ast}\bar{\mathrm{B}}^{\ast}(^{3}D_{2})$ & $\surd$ & $\surd$ \\ 
\hline
$2^{++}$ & $\mathrm{B}\bar{\mathrm{B}}(^{1}D_{2})$, $\frac{1}{\sqrt{2}} \left( \mathrm{B}\bar{\mathrm{B}}^{\ast}+\mathrm{B}^{\ast}\bar{\mathrm{B}} \right)(^{3}D_{2})$, $\mathrm{B}^{\ast}\bar{\mathrm{B}}^{\ast}(^{1}D_{2})$, $\mathrm{B}^{\ast}\bar{\mathrm{B}}^{\ast}(^{5}S_{2})$, $\mathrm{B}^{\ast}\bar{\mathrm{B}}^{\ast}(^{5}D_{2})$, $\mathrm{B}^{\ast}\bar{\mathrm{B}}^{\ast}(^{5}G_{2})$ & $\chi_{\mathrm{b}2}$ & $\surd$ \\ 
\hline
$2^{-+}$ & $\frac{1}{\sqrt{2}} \left( \mathrm{B}\bar{\mathrm{B}}^{\ast}-\mathrm{B}^{\ast}\bar{\mathrm{B}} \right)(^{3}P_{2})$, $\frac{1}{\sqrt{2}} \left( \mathrm{B}\bar{\mathrm{B}}^{\ast}-\mathrm{B}^{\ast}\bar{\mathrm{B}} \right)(^{3}F_{2})$, $\mathrm{B}^{\ast}\bar{\mathrm{B}}^{\ast}(^{3}P_{2})$, $\mathrm{B}^{\ast}\bar{\mathrm{B}}^{\ast}(^{3}F_{2})$ & $\eta_{\mathrm{b}2}$ & $\surd$ \\ 
\hline
$2^{--}$ & $\frac{1}{\sqrt{2}} \left( \mathrm{B}\bar{\mathrm{B}}^{\ast}+\mathrm{B}^{\ast}\bar{\mathrm{B}} \right)(^{3}P_{2})$, $\frac{1}{\sqrt{2}} \left( \mathrm{B}\bar{\mathrm{B}}^{\ast}+\mathrm{B}^{\ast}\bar{\mathrm{B}} \right)(^{3}F_{2})$, $\mathrm{B}^{\ast}\bar{\mathrm{B}}^{\ast}(^{5}P_{2})$, $\mathrm{B}^{\ast}\bar{\mathrm{B}}^{\ast}(^{5}F_{2})$ & $\psi_{\mathrm{b}2}$ & $\surd$ \\ 
\hline
\end{tabular}
}
\end{center}
\label{tbl:classification}
\end{table}%

We classify all the possible quantum numbers $I^{G}(J^{PC})$ with isospin $I$, $G$-parity, total angular momentum $J$, parity $P$ and charge conjugation $C$ for the states which can be composed by a pair of $\mathrm{B}^{(\ast)}$ and $\bar{\mathrm{B}}^{(\ast)}$ mesons.
The charge conjugation $C$ is defined for $I=0$
or $I_{z}=0$ components
for $I=1$, and is related to the $G$-parity by $G=(-1)^{I}C$.
In the present discussion, we restrict upper limit of the total angular
momentum as $J \le 2$, because too higher angular momentum
will be disfavored to form bound or resonant states.
The $\mathrm{B}^{(\ast)}\bar{\mathrm{B}}^{(\ast)}$ components 
in the wave functions for various $J^{PC}$ are 
listed in Table~\ref{tbl:classification}. 
We use the notation $^{2S+1}L_{J}$ to denote the total spin $S$ and
relative angular momentum $L$ of the two body states of $\mathrm{B}^{(\ast)}$ and $\bar{\mathrm{B}}^{(\ast)}$ mesons.
We note that there are not only $\mathrm{B}\bar{\mathrm{B}}$ and $\mathrm{B}^{\ast}\bar{\mathrm{B}}^{\ast}$ components but also $\mathrm{B}\bar{\mathrm{B}}^{\ast}\pm\bar{\mathrm{B}}\mathrm{B}^{\ast}$ components.
The $J^{PC} = 0^{+-}$ state cannot be generated by a combination of $\mathrm{B}^{(\ast)}$ and $\bar{\mathrm{B}}^{(\ast)}$ mesons \footnote{The $J^{PC} = 0^{+-}$ state cannot be given also in the quarkonium picture.}.
For $I=0$, there are many $\mathrm{B}^{(\ast)}\bar{\mathrm{B}}^{(\ast)}$
states whose quantum number $J^{PC}$ are the same as those of the quarkonia as shown
in the third row of $I=0$. 
In the present study, however, we do not consider these states,
because we have not yet included  mixing terms between the quarkonia
and the $\mathrm{B}^{(\ast)}\bar{\mathrm{B}}^{(\ast)}$ states. 
This problem will be left as future works.
Therefore, for $I=0$, we consider only the exotic
quantum numbers $J^{PC}=0^{--}$, $1^{-+}$ and $2^{+-}$.
The states of $I=1$ are
clearly not accessible by quarkonia. 
We investigate all possible 
 $J^{PC}$ states listed in Table~\ref{tbl:classification}.

From Eqs.~(\ref{eq:pot_BB*B*B})-(\ref{eq:pot_BB*B*B*}) and (\ref{eq:rhopot_BBBB})-(\ref{eq:rhopot_BB*B*B*}), we obtain the
potentials with channel-couplings for each quantum number $I^{G}(J^{PC})$.
For each state, the Hamiltonian is given as a sum of the kinetic energy and the  potential with channel-couplings in a form of a matrix.
Breaking of the heavy quark symmetry is taken into account by mass difference between $\mathrm{B}$ and $\mathrm{B}^{\ast}$ mesons in the kinetic term.
The explicit forms of the Hamiltonian for each $I^{G}(J^{PC})$ are presented in Appendix~\ref{sec:Hamiltonian}.
For example, the $J^{PC} = 1^{+-}$ state has four components,
$\frac{1}{\sqrt{2}} \left(
\mathrm{B}\bar{\mathrm{B}}^{\ast}-\mathrm{B}^{\ast}\bar{\mathrm{B}}
\right) (^{3}S_{1})$, $\frac{1}{\sqrt{2}} \left(
\mathrm{B}\bar{\mathrm{B}}^{\ast}-\mathrm{B}^{\ast}\bar{\mathrm{B}}
\right) (^{3}D_{1})$,
$\mathrm{B}^{\ast}\bar{\mathrm{B}}^{\ast}(^{3}S_{1})$,
$\mathrm{B}^{\ast}\bar{\mathrm{B}}^{\ast}(^{3}D_{1})$ and hence it
gives a  potential in the form of $4\times 4$ matrix as Eqs.~(\ref{eq:K_{1^{+-}}}), (\ref{eq:pi_pot_1+-}) and (\ref{eq:V^{v}_{1^{+-}}}).

\section{Numerical results}

To obtain the solutions of the $\mathrm{B}^{(\ast)}\bar{\mathrm{B}}^{(\ast)}$  states,
we solve numerically the Schr\"odinger equations which are  
second-order differential equations with channel-couplings.
As numerics, the renormalized Numerov method developed in
Ref.~\cite{johnson} is adopted.
The resonant states are found from the phase shift $\delta$ as a function of the scattering energy $E$.
The resonance position $E_r$ is defined by an inflection point of the 
phase shift $\delta(E)$ and the resonance width by 
$\Gamma_r = 2/(d\delta /dE)_{E=E_r}$  following Ref.~\cite{Arai:1999pg}.
To check consistency of our method with others,
we also use the complex scaling method (CSM) \cite{CSM}. We obtain an agreement 
in results between the renormalized Nemerov method and the CSM.

\begin{figure}[htbp]
\includegraphics[width=13cm]{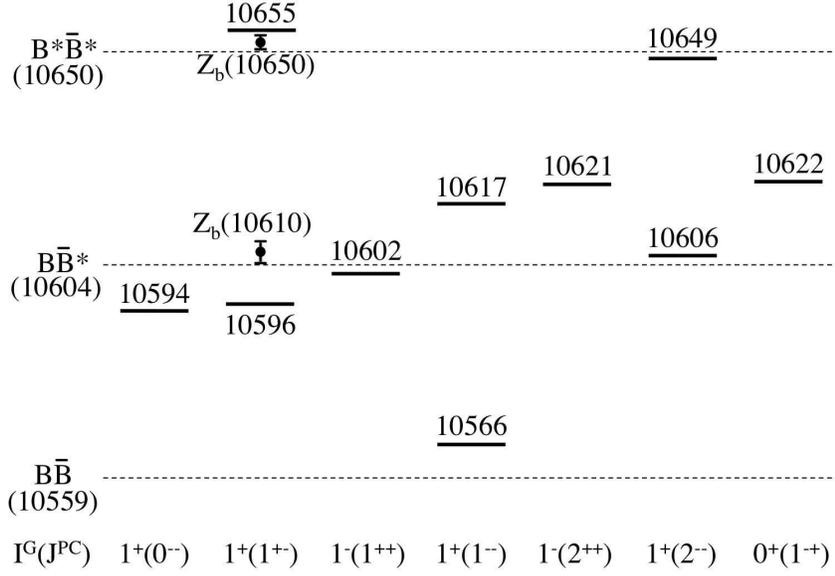}
\caption{The $\mathrm{B}^{(*)}\bar{\mathrm{B}}^{(*)}$ bound and
 resonant states with exotic $I^{G}(J^{PC})$. The dots with error bars denote the
 position of the experimentaly observed 
 $\mathrm{Z}_{\mathrm{b}}$'s where $M(\mathrm{Z}_{\mathrm{b}}(10610)) =
 10607.2$ MeV and $M (\mathrm{Z}_{\mathrm{b}}(10650)) = 10652.2$ MeV. Solid lines are for our predictions for the
 energies of the bound and resonant states when the $\pi \rho\, \omega$
 potential is employed. Mass values are shown in units of MeV.}
\label{fig:result}
\end{figure}

In Table~\ref{tbl:result_table}, we summarize the result of the obtained
 bound and resonant states, and their possible decay 
modes to quarkonium and light flavor meson.
For decay modes, the $\rho$ meson can be either real or virtual
 depending on the mass of the decaying particle, depending on the resonance energy which is either sufficient or not to emit the real state of $\rho$ or $\omega$ meson.
   $\rho^{\ast}(\omega^{\ast})$ indicates that it is a virtual state in radiative decays assuming the vector meson dominance.   
We show  the mass spectrum of these states in Fig~\ref{fig:result}.  

Let us see the states of isospin $I=1$.  
Interestingly, having the present potential we find the twin states in the $I^{G}(J^{PC})=1^{+}(1^{+-})$ near the
$\mathrm{B}\bar{\mathrm{B}}^{\ast}$ and  $\mathrm{B}^{\ast}\bar{\mathrm{B}}^{\ast}$
thresholds;  a bound state slightly below the $\mathrm{B}\bar{\mathrm{B}}^{\ast}$ threshold, and  a resonant state slightly above the $\mathrm{B}^{\ast}\bar{\mathrm{B}}^{\ast}$ threshold.
The binding energy is 8.5 MeV, and the resonance energy and decay width
are 50.4 MeV and 15.1 MeV, respectively, from the $\mathrm{B}\bar{\mathrm{B}}^{\ast}$ threshold.
The twin states are obtained when the $\pi \rho\, \omega$ potential is
used.  We interpret them as the
$\mathrm{Z}_\mathrm{b}(10610)$ and $\mathrm{Z}_\mathrm{b}(10650)$  observed in the Belle
experiment~\cite{Collaboration:2011gja,Belle:2011aa}.
It should be emphasized that the interaction in the present study has
been determined  in the previous works without knowing the experimental
data of $\mathrm{Z}_{\mathrm{b}}$'s \cite{Yasui:2009bz,Yamaguchi:2011xb}.
 
Several comments are in order.  
First, the bound state of lower energy has been obtained in the coupled
channel method of $\mathrm{B}\bar{\mathrm{B}}^{\ast}$ and
$\mathrm{B}^{\ast}\bar{\mathrm{B}}^{\ast}$  channels.  
In reality, however, they also couple to other lower channels such
as $\pi h_b$,  $\pi \Upsilon$ and so on as shown in Table~\ref{tbl:classification}.
Once these decay channels are included, the bound state will be a resonant 
state with a finite width.  
A qualitative discussion will be given in Section~5.
Second, when the $\pi$ exchange potential is used, only the lower bound state is
obtained but the resonant state is not.  However, we have verified
that a small change in the $\pi$ exchange potential generates, as well as the
bound state, the corresponding resonant state also.   
Therefore, the pion dominance is working for the  $\mathrm{B}\bar{\mathrm{B}}^{\ast}$ and $\mathrm{B}^{\ast}\bar{\mathrm{B}}^{\ast}$ systems. (See also the discussion in Appendix~\ref{sec:OPEP}.)
Third, it would provide a direct evidence of these states to be $\mathrm{B}\bar{\mathrm{B}}^{\ast}$ and $\mathrm{B}^{\ast}\bar{\mathrm{B}}^{\ast}$ molecules if the  $\mathrm{B}\bar{\mathrm{B}}^{\ast}$ and $\mathrm{B}^{\ast}\bar{\mathrm{B}}^{\ast}$ decays are observed in experiments.  
Whether the energies are below or above the thresholds is also checked by
the observation of these decays.  

In other channels, we  further predict the $\mathrm{B}^{(\ast)}\bar{\mathrm{B}}^{(\ast)}$ bound 
and resonant states.
The $I^{G}(J^{PC})=1^{-}(0^{++})$ state is a bound state with binding energy 6.5 MeV
from the $\mathrm{B}\bar{\mathrm{B}}$ threshold for the $\pi$ exchange potential, while no structure for the $\pi \rho\, \omega$ potential.
The existence of this state  depends on the
details of the potential, 
while the states in the other quantum numbers are rather robust.
Let us see the results for the latter states from the $\pi \rho\, \omega$ potentials. 
For $1^{+}(0^{--})$ and $1^{-}(1^{++})$, we find
bound states with binding energy 9.8 MeV and 1.9 MeV from the
$\mathrm{B}\bar{\mathrm{B}}^{\ast}$ threshold, respectively.
These bound states appear also for the $\pi$ exchange potential, though the binding energy of the $1^{-}(1^{++})$ state becomes larger.
The $1^{-}(2^{++})$ state is a resonant state with the resonance energy 62.7
MeV and the decay width 8.4 MeV.
The $1^{+}(1^{--})$ states are  twin resonances with the resonance energy 7.1
MeV and the decay width 37.4 MeV for the first resonance, and the
resonance energy 58.6 MeV and the decay width 27.7 MeV for the second.  
The resonance energies are measured from the $\mathrm{B}\bar{\mathrm{B}}$ threshold.
The $1^{+}(2^{--})$ states also form twin resonances with the resonance energy
2.0 MeV and the decay width 3.9 MeV for the first resonance and the
resonance energy 44.1 MeV and the decay width 2.8 MeV for the second, 
where the resonance energies have are  measured from the $\mathrm{B}\bar{\mathrm{B}}^{\ast}$ threshold.

Next we discuss the result for the states of isospin $I=0$.  
In general, the interaction in these states are either repulsive or 
only weakly attractive as compared to the cases of $I=1$.  
 The fact that there are less channel-couplings explains less attraction
 partly. (See also Appendix~\ref{sec:OPEP}.)
Because of this, we find only one resonant state with $I^{G}(J^{PC})=0^{+}(1^{-+})$, 
as shown in Fig~\ref{fig:result} and in Table~\ref{tbl:result_table_2}.  
The $0^{+}(1^{-+})$ state is a resonant state with the resonance energy 17.8
MeV and the decay width 30.1 MeV for the $\pi \rho\, \omega$ potential.

In the present study, all the states appear in the threshold regions and
therefore  are all weakly bound or
resonant states.
The present results are consequences of unique features of the bottom quark sector; the large reduced mass of the $\mathrm{B}^{(\ast)}\bar{\mathrm{B}}^{(\ast)}$ systems and the strong tensor force induced by the mixing of $\mathrm{B}$ and $\mathrm{B}^{\ast}$ with small mass splitting.
In fact, in the charm sector, our model does not predict any bound or
resonant states in the region where we research numerically. Because the reduced mass is smaller and the mass splitting between $\mathrm{D}$ and $\mathrm{D}^{\ast}$ is larger.  

\begin{table}[htbp]
\caption{\small Various properties of the
 $\mathrm{B}^{(\ast)}\bar{\mathrm{B}}^{(\ast)}$ bound and resonant
 states with possible $I^{G}(J^{PC})$ in $I=1$. The energies $E$ can be
 either pure real for bound states or complex for resonances.  The real
 parts are measured from the thresholds as indicated in the second
 column.  The imaginary parts are half of the decay widths of the
 resonances, $\Gamma/2$. In the last two columns, decay channels of a
 quarkonium and a light flavor meson are indicated. Asterisk of
 $\rho^{\ast}$ indicates that the decay occures only with a virtual
 $\rho$ while subsequently transit to a real photon via vector meson dominance.}  
\begin{center}
{
\begin{tabular}{|c|c|c|c|c|c|}
\hline
$I^G (J^{PC})$ & threshold &  \multicolumn{2}{c|}{$E$ [MeV]} & \multicolumn{2}{c|}{decay channels} \\
\cline{3-6}
 & & $\pi$-potential & $\pi \rho\, \omega$-potential & s-wave & p-wave \\
\hline
$1^+ (0^{+-})$ & --- & --- & --- &  --- & $\mathrm{h}_{\mathrm{b}}+\pi$, $\chi_{\mathrm{b}0,1,2} \!+\! \rho$ \\
\hline
$1^- (0^{++})$ & $\mathrm{B}\bar{\mathrm{B}}$ & $-6.5$  & no & $\eta_{\mathrm{b}} \!+\! \pi$, $\Upsilon \!+\! \rho$ & $\mathrm{h}_{\mathrm{b}} \!+\! \rho^{\ast}$, $\chi_{\mathrm{b}1} \!+\! \pi$ \\
\hline
$1^+ (0^{--})$ & $\mathrm{B}\bar{\mathrm{B}}^{\ast}$ & $-9.9$ & $-9.8$ & $\chi_{\mathrm{b}1} \!+\! \rho^{\ast}$ & $\eta_{\mathrm{b}} \!+\! \rho$, $\Upsilon \!+\! \pi$ \\
\hline
$1^- (0^{-+})$ & $\mathrm{B}\bar{\mathrm{B}}^{\ast}$ & no & no & $\mathrm{h}_{\mathrm{b}} \!+\! \rho$, $\chi_{\mathrm{b}0} \!+\! \pi$ & $\Upsilon \!+\! \rho$ \\
\hline
\multirow{2}*{$1^+ (1^{+-})$} & \multirow{2}*{$\mathrm{B}\bar{\mathrm{B}}^{\ast}$} & 
 \multirow{2}*{$-7.7$} & $-8.5$  &\multirow{2}*{ $\Upsilon \!+\! \pi$} &
 \multirow{2}*{$\mathrm{h}_{\mathrm{b}} \!+\! \pi$, $\chi_{\mathrm{b}1}
 \!+\! \rho^{\ast}$}  \\
 & &  & $50.4-i15.1/2$ & &\\
\hline
$1^- (1^{++})$ & $\mathrm{B}\bar{\mathrm{B}}^{\ast}$ & $-16.7$  & $-1.9$ & $\Upsilon \!+\! \rho$ & $\mathrm{h}_{\mathrm{b}} \!+\! \rho^{\ast}$, $\chi_{\mathrm{b}0,1} \!+\! \pi$ \\
\hline
\multirow{2}*{$1^+ (1^{--})$} & \multirow{2}*{$\mathrm{B}\bar{\mathrm{B}}$} & 
 $7.0-i37.9/2$ & $7.1-i37.4/2$ &\multirow{2}*{ $\mathrm{h}_{\mathrm{b}} \!+\! \pi$,
 $\chi_{\mathrm{b}0,1,2} \!+\! \rho^{\ast}$} & \multirow{2}*{
 $\eta_{\mathrm{b}} \!+\! \rho$, $\Upsilon \!+\! \pi$} \\
 & & $58.8-i30.0/2$ & $58.6-i27.7/2$ & &\\
\hline
$1^- (1^{-+})$ & $\mathrm{B}\bar{\mathrm{B}}^{\ast}$ & no  & no & $\mathrm{h}_{\mathrm{b}} \!+\! \rho$, $\chi_{\mathrm{b}1} \!+\! \pi$ & $\eta_{\mathrm{b}} \!+\! \pi$, $\Upsilon \!+\! \rho$ \\
\hline
$1^+ (2^{+-})$ & $\mathrm{B}\bar{\mathrm{B}}^{\ast}$ & no & no & --- & $\mathrm{h}_{\mathrm{b}} \!+\! \pi$, $\chi_{\mathrm{b}0,1,2} \!+\! \rho$ \\
\hline
$1^- (2^{++})$ & $\mathrm{B}\bar{\mathrm{B}}$ & $63.5-i8.3/2$ & $62.7-i8.4/2$ & $\Upsilon \!+\! \rho$ & $\mathrm{h}_{\mathrm{b}} \!+\! \rho^{\ast}$, $\chi_{\mathrm{b}1,2} \!+\! \pi$ \\
\hline
$1^- (2^{-+})$ & $\mathrm{B}\bar{\mathrm{B}}^{\ast}$ & no & no & $\mathrm{h}_{\mathrm{b}} \!+\! \rho$ & $\Upsilon \!+\! \rho$ \\
\hline
\multirow{2}*{$1^+ (2^{--})$} & \multirow{2}*{$\mathrm{B}\bar{\mathrm{B}}^{\ast}$} & 
 $2.0-i4.1/2$ & $2.0-i3.9/2$ &\multirow{2}*{$\chi_{\mathrm{b}1} \!+\! \rho^{\ast}$} &
 \multirow{2}*{ $\eta_{\mathrm{b}} \!+\! \rho$, $\Upsilon \!+\! \pi$}  \\
 & & $44.2-i2.5/2$ & $44.1-i2.8/2$ & & \\
\hline
\end{tabular}
}
\end{center}
\label{tbl:result_table}
\end{table}%

\begin{table}[htbp]
\caption{\small The $\mathrm{B}^{(\ast)}\bar{\mathrm{B}}^{(\ast)}$ bound and resonant states with exotic $I^{G}(J^{PC})$ in $I=0$. (Same convention as Table II.)}
\begin{center}
{\renewcommand\arraystretch{1.2}
\begin{tabular}{|c|c|c|c|c|c|}
\hline
$I^G (J^{PC})$ & threshold &  \multicolumn{2}{c|}{$E$ [MeV]} & \multicolumn{2}{c|}{decay channels} \\
\cline{3-6}
 & & $\pi$-potential & $\pi \rho\, \omega$-potential & s-wave & p-wave \\
\hline
$0^- (0^{--})$ & $\mathrm{B}\bar{\mathrm{B}}^{\ast}$ & no & no
	     &$\chi_{\mathrm{b}1} \!+\! \omega$ &  $\eta_{\mathrm{b}} \!+\! \omega$, $\Upsilon \!+\! \eta$ \\
\hline
$0^+ (1^{-+})$ & $\mathrm{B}\bar{\mathrm{B}}^{\ast}$ & $28.6-i91.6/2$ & $17.8-i30.1/2$
	     &$\mathrm{h}_{\mathrm{b}} \!+\! \omega^{\ast}$, $\chi_{\mathrm{b}1}
 \!+\! \eta$ & $\eta_{\mathrm{b}} \!+\! \eta$, $\Upsilon \!+\! \omega$ \\
\hline
$0^- (2^{+-})$ & $\mathrm{B}\bar{\mathrm{B}}^{\ast}$ & no & no  & --- &
 $\mathrm{h}_{\mathrm{b}} \!+\! \eta$, $\chi_{\mathrm{b}0,1,2} \!+\! \omega$ \\
\hline
\end{tabular}
}
\end{center}
\label{tbl:result_table_2}
\end{table}%

\section{Effects of the coupling to decay channels}

We have employed the hadronic molecular picture and only considered 
the $\mathrm{B}^{(\ast)}\bar{\mathrm{B}}^{(\ast)}$ states so far.
In reality, however, the $\mathrm{B}^{(\ast)}\bar{\mathrm{B}}^{(\ast)}$ states
couple to a bottomonium and a light meson state which is predominantly
a pion, as $\mathrm{Z}_{\mathrm{b}}$'s were discovered  in the decay channels of $\Upsilon (nS)\pi$ ($n=1,2,3$)
and ${\mathrm h}_{\mathrm b}(mP)\pi$ ($m=1,2$)~\cite{Collaboration:2011gja,Belle:2011aa}.
In this section, we estimate the effects of such channel coupling to
the $\mathrm{B}^{(*)}\bar{\mathrm{B}}^{(*)}$ states.
We give  a qualitative estimation for the lowest ${\mathrm B}^{(\ast)}\bar{\mathrm
B}^{(\ast)}$ state in $1^{+}(1^{+-})$  corresponding to $\mathrm{Z}_{\mathrm{b}}(10610)^{\pm}$.
Similar effects are expected for other states.

\begin{figure}[htbp]
\begin{center}
 \includegraphics{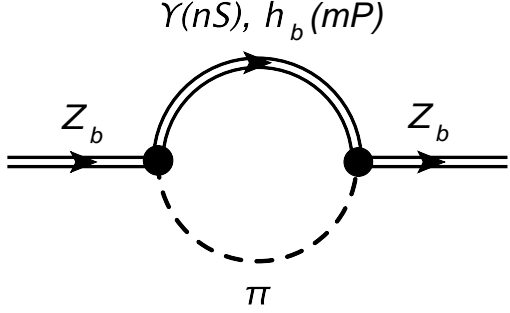}
 \caption{The diagram corresponding to a loop function $\Pi_n (s)$ of channel $n$. }
\label{loopdiag}
\end{center}
\end{figure}

To this purpose, we employ the method of Pennington and Wilson~\cite{Pennington:2007xr}.
They calculated charmonium mass-shifts for including the effect of open
and nearby closed channels and we apply their calculation procedure 
for $\mathrm{Z}_{\mathrm{b}}$ mass-shift.
The bare bound state propagator $i/[s-m^2_0]$, where $m_0$ is the mass
of the bare state, is dressed by the
contribution of hadron loops $\Pi (s)$. Therefore, the full propagator
can be written as 
\begin{eqnarray}
 G_z (s) = \frac{i}{s- \mathcal{M}^2 (s)}&=& \frac{i}{s- m^2_0 - \Pi (s)} \\ \notag
         &=& \frac{i}{s- m^2_0 - \sum_{n=1} \Pi_n (s)}  \, , 
\end{eqnarray}
where $s$ is the square of the momentum carried by the propagator.
$\mathcal{M}(s)$ is the complex mass function and the real part of this
give the ``renormalized'' mass.
Since the $\mathrm{Z}_{\mathrm{b}}$ has five decay channels,
the hadron loops $\Pi (s)$ is a sum of each decay channel $n$~(Fig.\ref{loopdiag}).
Each hadron loop $\Pi_n(s)$ (Fig.\ref{loopdiag}) is obtained by using the dispersion relation
in terms of its imaginary part.
All hadronic channels contribute to its mass at least in
principle.
Because the dispersion integral diverges, we have to subtract the square 
of mass function $\mathcal{M}(s_0)$ at suitable point $s_0$ from
$\mathcal{M}(s)$.
We shall discuss the choice of $s_0$ shortly.
Now, we can write the loop function in a once subtracted form as
\begin{eqnarray}
\Delta \Pi_n (s,s_0) \equiv \Pi_n (s) - \Pi_n (s_0)
= \frac{(s- s_0 )}{\pi} \int^{\infty}_{s_n} ds^{\prime}
\frac{\mathrm{Im} \Pi_n (s^{\prime})}{(s^{\prime} - s) (s^{\prime} -
s_0)} \, .
\label{dispersion}
\end{eqnarray}
Then we arrive at the mass-shift $\delta M$ as
\begin{eqnarray}
 \sum_{n=1}\Delta \Pi_n (s,s_0) = \mathcal{M}^2 (s) -m_0^2 \equiv \delta M^2 (s) \, .
\end{eqnarray}
Since an imaginary part of a loop function is proportional to  the
two-body phase space,
we take $\mathrm{Im} \Pi_n$ in the form for $s \geq s_n$ as
\begin{eqnarray}
 \mathrm{Im}\Pi_n (s) = -g^2_n \left( \frac{2
				q_{cm}}{\sqrt{s}}\right)^{2L+1}
 \exp{\left(-\frac{q_{cm}^2}{{\Lambda}^2}\right)} \, ,
\label{phasespace}
\end{eqnarray}
where $g_n$ is the coupling of $\mathrm{Z}_{\mathrm{b}}$ to a decay
channel $n$ (a bottomonium and a pion), $L$ is the orbital angular momentum
between a bottomonium and a pion. $q_{cm}$ is the magnitude of the three
momentum of a pion in the center of mass frame and is related to $q_{cm}$ by
\begin{eqnarray}
 q_{cm}=\left( \frac{s +m^2_{\pi} -
	 M^2_{b\bar{b}}}{2\sqrt{s}}\right)^2 -m^2_{\pi} \, .
\end{eqnarray}
In eq~(\ref{phasespace}), following~\cite{Pennington:2007xr}, we have introduced the Gaussian-type form
factor with a  cut-off parameter $\Lambda$ 
which is related to the interaction range $R$.
We set $\Lambda = 600$ MeV as a typical hadron scale; this value corresponds 
to $R \sim 0.8$ fm by using the relation $R \simeq \sqrt{6}/ \Lambda $.
Coupling $g_n$ is determined from the partial decay width $\Gamma_n$, 
by $\Gamma_n (s) = - \mathrm{Im}\Pi_n (s)/ \sqrt{s}$.
For the present rough estimation, we postulate that the decay rates for five final states ($\Upsilon (1S)
\pi$,  $\Upsilon (2S) \pi$,  $\Upsilon (3S) \pi$,  $h_b (1P) \pi$,
$h_b (2P) \pi $) are equal.
Then partial decay width for each decay channel is set as 3 MeV that 
is one-fifth of the total decay width 15 MeV~\cite{Collaboration:2011gja,Belle:2011aa}.

The subtraction point $s_0$ determines the renormalization point 
where the loop correction vanishes.
In Ref.~\cite{Pennington:2007xr}, the subtraction point was chosen at 
the mass of $J/\psi$.
Since $J/\psi$ is a deeply bound state of a $c\bar{c}$ pair where the
charmonium discription works well without a $\mathrm{D}\mathrm{\bar{D}}$
loop.
Now in our situation, there is no such a physical bottomonium like
state decaying into a pion and a bottomonium. 
However, as in the case of $J/\psi$ we expect that the renormalization
point of the vanishing loop is located at an energy which is
significantly below the thresholods of the particle in the loop.
We adopt such an energy at $\sqrt{s_0} = 9000$ MeV, 600 MeV below the 
$\pi \Upsilon (1S)$, which is similar to the mass difference of 
$J/\psi$ and $\mathrm{D}\mathrm{\bar{D}}$.

The resulting mass-shift $\delta M$ due to each coupling is given in Table~\ref{mass-shift}.
The total mass-shift is $\delta M = 2.4$ MeV, which is slightly
repulsive.
This means that the mass of the ${\mathrm B}^{(\ast)}\bar{\mathrm B}^{(\ast)}$ bound state in $1^{+}(1^{+-})$ will be pushed up by the $\Upsilon(nS)\pi$ and ${\mathrm h}_{\mathrm b}(mP)\pi$ couplings.
Therefore, we expect that this state gets closer the ${\mathrm B}\bar{\mathrm B}^{\ast}$ threshold, or could even become a resonant state.
Since the coupling $g_{\Upsilon (1S)\pi}$  is the largest due to its low mass,
 the largest effect is found for the coupling of $\Upsilon(1S)\pi$,
where the mass-shift $\delta M$ is 6.3 MeV.
The coupling of $h_b (2P)\pi$ having $P$-wave contributes attraction,
whose mass-shift $\delta M$ is -3.0 MeV.
Other coupling channels are minor role. 

To summarize this section, we have estimated loop contributions to the
mass of the  ${\mathrm B}^{(\ast)}\bar{\mathrm B}^{(\ast)}$ molecules.
We find small repulsive corrections, which still keeps the molecular
picture unchanged but may change the bound states into resonances, being
consistent with the experimental observation.

\begin{table}[htbp]
 \centering
 \caption{\label{table:massshift} \small Various contributions to loop corrections of channel $n$, $\delta M$.
The total correction is shown on the most right column. The first and
 the second row show the threshold masses and the coupling strenghts of
 channel $n$. $M_{th}$ and $\delta M$ are given in units of MeV.}
\begin{tabular}{c | c c c c c c}
 & $\Upsilon (1S)\pi$ & $\Upsilon (2S)\pi$ & $\Upsilon (3S)\pi$ &
 ${\mathrm h}_{\mathrm b} (1P)\pi$ & ${\mathrm h}_{\mathrm b} (2P)\pi$ &
 total \\ 
\hline
 $M_{th}$  & 9600 & 10163 & 10495 & 10038 & 10399 & --- \\
 $g_n$ & 1986 & 844 & 956 & 7392 & 14179 & --- \\
 $\delta M$ & 6.3 & 0.5 & -1.3 & -0.1 & -3.0 & 2.4 \\
\end{tabular}
\label{mass-shift}
\end{table}

\section{Search in decays from $\Upsilon(5S)$}

As  twin $\mathrm{Z}_{\mathrm{b}}$'s were observed from
$\Upsilon(5\mathrm{S})$ decay, $\Upsilon(5\mathrm{S})$ decay is a useful
source to search the exotic states around the
$\mathrm{B}^{(\ast)}\bar{\mathrm{B}}^{(\ast)}$ energy region.
$\Upsilon(5\mathrm{S})$ can decay to a $I^{G}(J^{PC})=1^{+}(1^{+-})$ state by a single pion emission in s-wave, and a $1^{+}(0^{--})$, $1^{+}(1^{--})$ or $1^{+}(2^{--})$ state by a single pion emission in p-wave.
We recall that the twin $\mathrm{Z}_{\mathrm{b}}$'s with $I^{G}(J^{PC})=1^{+}(1^{+-})$ were observed in the s-wave channel~\cite{Collaboration:2011gja,Belle:2011aa}.
In the present study, we further predict the bound state in $I^{G}(J^{PC})=1^{+}(0^{--})$, and another twin resonant states in $I^{G}(J^{PC})=1^{+}(1^{--})$ and $1^{+}(2^{--})$ as summarized in Table~\ref{tbl:result_table}.
As for the exotic $J^{PC}$ states in isosinglet, the resonant state in $I^{G}(J^{PC})=0^{+}(1^{-+})$ can be observed from $\Upsilon(5S)$ by $\omega$ emission in p-wave as shown in Table~\ref{tbl:result_table_2}.

The radiative decay of $\Upsilon(5S)$ is also an interesting channel as discussed in Ref.~\cite{Voloshin:2011qa}.
In radiative decay, $\Upsilon$(5S) decays to the $I^{G}(J^{PC})=1^-(0^{++})$, $1^-(1^{++})$ and $1^-(2^{++})$ states with a photon emission in s-wave.
These channels can be also produced in hadronic transitions with emission
of $\rho$ meson from higher $\Upsilon$-like bottomonim states.
In the present study, we predict the bound states in $I^{G}(J^{PC})=1^{-}(0^{++})$ and $1^{-}(1^{++})$ and a resonant states in $1^{-}(2^{++})$ as summarized in Table~\ref{tbl:result_table}.

As a consequence, we will be able to study the $\mathrm{B}^{(\ast)}\bar{\mathrm{B}}^{(\ast)}$ bound and resonant states with positive $G$-parity in a pion emission from $\Upsilon(5S)$ and with negative $G$-parity in a photon emission from $\Upsilon(5S)$.
It will be an interesting subject for experiments to search these states in $\Upsilon(5S)$ decays.

\section{Summary}

In this paper, we have systematically studied the possibility of 
 the $\mathrm{B}^{(\ast)}\bar{\mathrm{B}}^{(\ast)}$
bound and resonant states having  exotic quantum numbers $I^{G}(J^{PC})$.
These states are consisted of at least four quarks, because
their quantum numbers cannot be assigned by the quarkonium picture and hence they are genuinely exotic states.
We have constructed the potential of the
 $\mathrm{B}^{(\ast)}\bar{\mathrm{B}}^{(\ast)}$ states using the effective
Lagrangian respecting the heavy quark symmetry.
Because of the degeneracy in masses of $\mathrm{B}$ and $\mathrm{B}^{\ast}$ mesons,
the channel mixing, such as $\mathrm{B} \bar{\mathrm{B}}^{\ast}$-$\mathrm{B}^{\ast} \bar{\mathrm{B}}$, $\mathrm{B}^{\ast} \bar{\mathrm{B}}^{\ast}$-$\mathrm{B}^{\ast} \bar{\mathrm{B}}^{\ast}$, $\mathrm{B} \bar{\mathrm{B}}$-$\mathrm{B}^{\ast} \bar{\mathrm{B}}^{\ast}$ and $\mathrm{B} \bar{\mathrm{B}}^{\ast}$-$\mathrm{B}^{\ast} \bar{\mathrm{B}}^{\ast}$, plays an important role to form the $\mathrm{B}^{(\ast)} \bar{\mathrm{B}}^{(\ast)}$ bound and/or resonant states.
We have numerically solved the Schr\"odinger equation with
the channel-couplings for the $\mathrm{B}^{(\ast)} \bar{\mathrm{B}}^{(\ast)}$ states with $I^{G}(J^{PC})$ for $J \le 2$.

As a result, in $I=1$, we have found that the
$I^{G}(J^{PC})=1^{+}(1^{+-})$ states have a bound state with binding 
energy 8.5 MeV, and a resonant state with the resonance energy 50.4 MeV 
and the decay width 15.1 MeV.
We have successfully reproduced the positions of
$\mathrm{Z}_{\mathrm b}(10610)$ and $\mathrm{Z}_{\mathrm b}(10650)$ observed by Belle.
Therefore, the twin resonances of $\mathrm{Z}_{\mathrm b}$'s can be interpreted as
the $\mathrm{B}^{({\ast})}\bar{\mathrm{B}}^{({\ast})}$ molecular type states.
It should be noted that the $\mathrm{B} \bar{\mathrm{B}}^{\ast}$-$\mathrm{B}^{\ast} \bar{\mathrm{B}}$, $\mathrm{B} \bar{\mathrm{B}}^{\ast}$-$\mathrm{B}^{\ast} \bar{\mathrm{B}}^{\ast}$ and $\mathrm{B}^{\ast} \bar{\mathrm{B}}$-$\mathrm{B}^{\ast} \bar{\mathrm{B}}^{\ast}$ mixing effects are important, because many structures disappear without the mixing effects.
We have obtained the other possible
$\mathrm{B}^{({\ast})}\bar{\mathrm{B}}^{({\ast})}$ states in $I=1$.
We have found  one bound state in  each $1^{+}(0^{--})$ and  $1^{-}(1^{++})$, one resonant state in $1^{-}(2^{++})$ and twin resonant states in each $1^{+}(1^{--})$ and $1^{+}(2^{--})$.
It is remarkable that another two twin resonances can exist in addition to the $\mathrm{Z}_{\mathrm{b}}$'s.
We have also studied the $\mathrm{B}^{({\ast})}\bar{\mathrm{B}}^{({\ast})}$ states in $I=0$ and found one resonant state in $0^{+}(1^{-+})$.
We have checked the differences between the results from the $\pi$
exchange potential and those from the $\pi \rho\, \omega$ potential, and found that the difference is small.
Therefore, the one pion exchange potential dominates as the interaction in the $\mathrm{B}^{(\ast)}\bar{\mathrm{B}}^{(\ast)}$ bound and resonant states.

We have estimated the effects of the coupling to decay channels by means
of dispersion relations.
Total mass-shift is $\delta M = 2.4$ MeV, which is slightly repulsive.
Therefore, we conclude that the molecular picture of $\mathrm{B}^{(\ast)}\bar{\mathrm{B}}^{(\ast)}$ will be a good approximation  for the first step.
More systematic analyses will be left for future works.

For experimental studies,
the $\Upsilon(5S)$ decay is a useful tool to search the 
$\mathrm{B}^{({\ast})}\bar{\mathrm{B}}^{({\ast})}$ states.
$\Upsilon(5S)$ can decay to the
$\mathrm{B}^{(\ast)}\bar{\mathrm{B}}^{(\ast)}$ states with $1^{+}(0^{--})$, $1^{+}(1^{--})$ and $1^{+}(2^{--})$ by a single pion emission in p-wave and the state with $0^{+}(1^{-+})$ by $\omega$ emission in p-wave.
$\Upsilon(5S)$ can also decay to the
$\mathrm{B}^{(\ast)}\bar{\mathrm{B}}^{(\ast)}$ states 
with $1^{-}(0^{++})$, $1^{-}(1^{++})$ and $1^{-}(2^{++})$ by radiative decays.
In the future, various exotic states would be observed around the
thresholds from $\Upsilon(5S)$ decays in accelerator facilities such as Belle and also would be searched in the relativistic heavy ion collisions in RHIC and LHC~\cite{Cho:2010db,Cho:2011ew}.
If these states are fit in our predictions, they  will be good
candidates of the $\mathrm{B}^{({\ast})}\bar{\mathrm{B}}^{({\ast})}$ molecular states.

\section*{Acknowledgments}
We thank  Prof.~S.~Takeuchi and Prof.~M.~Takizawa for fruitful discussions and comments.
This work is supported in part by Grant-in-Aid for Scientific Research on 
Priority Areas ``Elucidation of New Hadrons with a Variety of Flavors 
(E01: 21105006)" (S.Y. and A.H.) and by ``Grant-in-Aid for Young Scientists (B)
22740174" (K.S.), from 
the ministry of Education, Culture, Sports, Science and Technology of
Japan.

\appendix

\section{Hamiltonian} \label{sec:Hamiltonian}

The hamiltonian is a sum of the kinetic term and potential term as,
\begin{eqnarray}
H_{J^{PC}} = K_{J^{PC}} + V^{\pi}_{J^{PC}},
\end{eqnarray}
for the $\pi$ exchage potential only, and 
\begin{eqnarray}
H_{J^{PC}} = K_{J^{PC}} + \sum_{i=\pi, \rho, \omega}V^{i}_{J^{PC}},
\end{eqnarray}
for the $\pi \rho\, \omega$ potential.

The kinetic terms with including the explicit breaking of the heavy quark symmetry by the mass difference $m_{\mathrm{B}^{\ast}}-m_{\mathrm{B}}$ are
\begin{align}
K_{0^{++}} &=
\mathrm{diag} \left(
 -\frac{1}{2\tilde{m}_{\mathrm{B}\mathrm{B}}} \triangle_{0},
 -\frac{1}{2 \tilde{m}_{\mathrm{B}\mathrm{B}^{\ast}}} \triangle_{0} + 2 \Delta m_{\mathrm{B}\mathrm{B}^{\ast}},
 -\frac{1}{2\tilde{m}_{\mathrm{B}\mathrm{B}^{\ast}}} \triangle_{2} + 2 \Delta m_{\mathrm{B}\mathrm{B}^{\ast}}
  \right), \\
K_{0^{--}} &=
\mathrm{diag} \left(
 -\frac{1}{2\tilde{m}_{\mathrm{B}\mathrm{B}^{\ast}}} \triangle_{1}
  \right), \\
K_{0^{-+}} &=
\mathrm{diag} \left(
 -\frac{1}{2\tilde{m}_{\mathrm{B}\mathrm{B}^{\ast}}} \triangle_{1},
 -\frac{1}{2\tilde{m}_{\mathrm{B}^{\ast}\mathrm{B}^{\ast}}} \triangle_{1} + \Delta m_{\mathrm{B}\mathrm{B}^{\ast}}
  \right), \\
K_{1^{+-}} &=
\mathrm{diag} \left(
 -\frac{1}{2\tilde{m}_{\mathrm{B}\mathrm{B}^{\ast}}} \triangle_{0},
 -\frac{1}{2\tilde{m}_{\mathrm{B}\mathrm{B}^{\ast}}} \triangle_{2},
 -\frac{1}{2\tilde{m}_{\mathrm{B}^{\ast}\mathrm{B}^{\ast}}} \triangle_{0} + \Delta m_{\mathrm{B}\mathrm{B}^{\ast}},
 -\frac{1}{2\tilde{m}_{\mathrm{B}^{\ast}\mathrm{B}^{\ast}}} \triangle_{2} + \Delta m_{\mathrm{B}\mathrm{B}^{\ast}}
 \right), \label{eq:K_{1^{+-}}} \\
K_{1^{++}} &=
\mathrm{diag} \left(
 -\frac{1}{2\tilde{m}_{\mathrm{B}\mathrm{B}^{\ast}}} \triangle_{0},
 -\frac{1}{2\tilde{m}_{\mathrm{B}\mathrm{B}^{\ast}}} \triangle_{2},
 -\frac{1}{2\tilde{m}_{\mathrm{B}^{\ast}\mathrm{B}^{\ast}}} \triangle_{2} + \Delta m_{\mathrm{B}\mathrm{B}^{\ast}}
 \right), \\
K_{1^{--}} &=
\mathrm{diag} \left(
 -\frac{1}{2\tilde{m}_{\mathrm{B}\mathrm{B}}} \triangle_{0},
 -\frac{1}{2\tilde{m}_{\mathrm{B}\mathrm{B}^{\ast}}} \triangle_{1} + \Delta m_{\mathrm{B}\mathrm{B}^{\ast}}, \right. \nonumber \\ 
& \left. -\frac{1}{2\tilde{m}_{\mathrm{B}^{\ast}\mathrm{B}^{\ast}}} \triangle_{1} + 2\Delta m_{\mathrm{B}\mathrm{B}^{\ast}},
 -\frac{1}{2\tilde{m}_{\mathrm{B}^{\ast}\mathrm{B}^{\ast}}} \triangle_{1} + 2\Delta m_{\mathrm{B}\mathrm{B}^{\ast}},
 -\frac{1}{2\tilde{m}_{\mathrm{B}^{\ast}\mathrm{B}^{\ast}}} \triangle_{3} + 2\Delta m_{\mathrm{B}\mathrm{B}^{\ast}}
 \right), \\
K_{1^{-+}} &=
\mathrm{diag} \left(
 -\frac{1}{2\tilde{m}_{\mathrm{B}\mathrm{B}^{\ast}}} \triangle_{1},
 -\frac{1}{2\tilde{m}_{\mathrm{B}^{\ast}\mathrm{B}^{\ast}}} \triangle_{1} + \Delta m_{\mathrm{B}\mathrm{B}^{\ast}}
 \right), \\
K_{2^{+-}} &=
\mathrm{diag} \left(
 -\frac{1}{2\tilde{m}_{\mathrm{B}\mathrm{B}^{\ast}}} \triangle_{2},
 -\frac{1}{2\tilde{m}_{\mathrm{B}^{\ast}\mathrm{B}^{\ast}}} \triangle_{2} + \Delta m_{\mathrm{B}\mathrm{B}^{\ast}}
 \right), \\
K_{2^{++}} &=
\mathrm{diag} \left(
 -\frac{1}{2\tilde{m}_{\mathrm{B}\mathrm{B}}} \triangle_{2},
 -\frac{1}{2\tilde{m}_{\mathrm{B}\mathrm{B}^{\ast}}} \triangle_{2} + \Delta m_{\mathrm{B}\mathrm{B}^{\ast}},
 -\frac{1}{2\tilde{m}_{\mathrm{B}^{\ast}\mathrm{B}^{\ast}}} \triangle_{2} + 2 \Delta m_{\mathrm{B}\mathrm{B}^{\ast}}, \right. \nonumber \\
& \left. -\frac{1}{2\tilde{m}_{\mathrm{B}^{\ast}\mathrm{B}^{\ast}}} \triangle_{0} + 2 \Delta m_{\mathrm{B}\mathrm{B}^{\ast}},
 -\frac{1}{2\tilde{m}_{\mathrm{B}^{\ast}\mathrm{B}^{\ast}}} \triangle_{2} + 2 \Delta m_{\mathrm{B}\mathrm{B}^{\ast}},
 -\frac{1}{2\tilde{m}_{\mathrm{B}^{\ast}\mathrm{B}^{\ast}}} \triangle_{4} + 2 \Delta m_{\mathrm{B}\mathrm{B}^{\ast}}
 \right), \\
K_{2^{-+}} &=
\mathrm{diag} \left(
 -\frac{1}{2\tilde{m}_{\mathrm{B}\mathrm{B}^{\ast}}} \triangle_{1},
 -\frac{1}{2\tilde{m}_{\mathrm{B}\mathrm{B}^{\ast}}} \triangle_{3},
 -\frac{1}{2\tilde{m}_{\mathrm{B}^{\ast}\mathrm{B}^{\ast}}} \triangle_{1} + \Delta m_{\mathrm{B}\mathrm{B}^{\ast}},
 -\frac{1}{2\tilde{m}_{\mathrm{B}^{\ast}\mathrm{B}^{\ast}}} \triangle_{3} + \Delta m_{\mathrm{B}\mathrm{B}^{\ast}}
 \right), \\
K_{2^{--}} &=
\mathrm{diag} \left(
 -\frac{1}{2\tilde{m}_{\mathrm{B}\mathrm{B}^{\ast}}} \triangle_{1},
 -\frac{1}{2\tilde{m}_{\mathrm{B}\mathrm{B}^{\ast}}} \triangle_{3},
 -\frac{1}{2\tilde{m}_{\mathrm{B}^{\ast}\mathrm{B}^{\ast}}} \triangle_{1} + \Delta m_{\mathrm{B}\mathrm{B}^{\ast}},
 -\frac{1}{2\tilde{m}_{\mathrm{B}^{\ast}\mathrm{B}^{\ast}}} \triangle_{3} + \Delta m_{\mathrm{B}\mathrm{B}^{\ast}}
 \right),
\end{align}
where $\triangle_{l} = \frac{\partial^{2}}{\partial r^{2}}+\frac{2}{r} \frac{\partial}{\partial r} - \frac{l(l+1)}{r^{2}}$ with integer $l \ge 0$, $1/\tilde{m}_{\mathrm{B}\mathrm{B}} = 1/m_{\mathrm{B}}+1/m_{\mathrm{B}}$, $1/\tilde{m}_{\mathrm{B}\mathrm{B}^{\ast}} = 1/m_{\mathrm{B}}+1/m_{\mathrm{B}^{\ast}}$, $1/\tilde{m}_{\mathrm{B}^{\ast}\mathrm{B}^{\ast}} = 1/m_{\mathrm{B}^{\ast}}+1/m_{\mathrm{B}^{\ast}}$ and  $\Delta m_{\mathrm{B}\mathrm{B}^{\ast}} = m_{\mathrm{B}^{\ast}} - m_{\mathrm{B}}$.

\newpage
The $\pi$ exchange potentials for each $J^{PC}$ states are
{\scriptsize
\begin{eqnarray}
V_{0^{++}}^{\pi} &=&
\left(
 \begin{array}{ccc}
  0 & \sqrt{3} V_{\mathrm C} & -\sqrt{6} V_{\mathrm T}   \\
  \sqrt{3} V_{\mathrm C} & 2V_{\mathrm C}  & \sqrt{2} V_{\mathrm T}  \\
 -\sqrt{6} V_{\mathrm T} & \sqrt{2} V_{\mathrm T}  &  -V_{\mathrm C}+2V_{\mathrm T}
\end{array}
\right) \, , 
\label{eq:pi_pot_0++} \\
V_{0^{--}}^{\pi} &=&
\left(
-V_{\mathrm C}-2V_{\mathrm T}
\right) \, , \\
V_{0^{-+}}^{\pi} &=&
\left(
\begin{array}{cc}
  V_{\mathrm C}+2V_{\mathrm T} & -2V_{\mathrm C}+2V_{\mathrm T}    \\
  -2V_{\mathrm C}+2V_{\mathrm T} & V_{\mathrm C}+2V_{\mathrm T}
\end{array}
\right) \, ,
\label{eq:pi_pot_0-+} \\
V_{1^{+-}}^{\pi} &=&
\left(
\begin{array}{cccc}
  V_{\mathrm C} & -\sqrt{2} V_{\mathrm T} & -2V_{\mathrm C} & -\sqrt{2} V_{\mathrm T}   \\
  -\sqrt{2} V_{\mathrm T} & V_{\mathrm C}+V_{\mathrm T}  & -\sqrt{2} V_{\mathrm T} & -2V_{\mathrm C} + V_{\mathrm T} \\
 -2V_{\mathrm C} & -\sqrt{2} V_{\mathrm T}  & V_{\mathrm C} & -\sqrt{2} V_{\mathrm T} \\
 -\sqrt{2} V_{\mathrm T} & -2V_{\mathrm C} + V_{\mathrm T} & -\sqrt{2} V_{\mathrm T} & V_{\mathrm C} + V_{\mathrm T}
\end{array}
\right) \, ,
\label{eq:pi_pot_1+-} \\
V_{1^{++}}^{\pi} &=&
\left(
\begin{array}{ccc}
  -V_{\mathrm C} & \sqrt{2} V_{\mathrm T} & \sqrt{6} V_{\mathrm T}   \\
  \sqrt{2} V_{\mathrm T} & -V_{\mathrm C}-V_{\mathrm T} & \sqrt{3} V_{\mathrm T}  \\
 \sqrt{6} V_{\mathrm T} & \sqrt{3} V_{\mathrm T} &  -V_{\mathrm C}+V_{\mathrm T}
\end{array}
\right) \, ,
\label{eq:pi_pot_1++} \\
V_{1^{--}}^{\pi} &=&
\left(
\begin{array}{ccccc}
  0 & 0 & \sqrt{3} V_{\mathrm C} & 2\sqrt{\frac{3}{5}} V_{\mathrm T} & -3\sqrt{\frac{2}{5}} V_{\mathrm T}   \\
  0 & -V_{\mathrm C}+V_{\mathrm T} & 0  & 3\sqrt{\frac{3}{5}} V_{\mathrm T} & 3\sqrt{\frac{2}{5}} V_{\mathrm T} \\
 \sqrt{3} V_{\mathrm C} & 0 & 2 V_{\mathrm C}  & -\frac{2}{\sqrt{5}}V_{\mathrm C} & \sqrt{\frac{6}{5}} V_{\mathrm T} \\
  2\sqrt{\frac{3}{5}} V_{\mathrm T} & 3\sqrt{\frac{2}{5}} V_{\mathrm T} & -\frac{2}{\sqrt{5}}V_{\mathrm C} & -V_{\mathrm C} + \frac{7}{5} V_{\mathrm T} & -\frac{\sqrt{6}}{5} V_{\mathrm T} \\
  -3\sqrt{\frac{2}{5}} V_{\mathrm T} & 3\sqrt{\frac{2}{5}} V_{\mathrm T} & \sqrt{\frac{6}{5}} V_{\mathrm T} & -\frac{\sqrt{6}}{5} V_{\mathrm T} & -V_{\mathrm C} + \frac{8}{5} V_{\mathrm T}
\end{array}
\right) \, ,
\label{eq:pi_pot_1--} \\
V_{1^{-+}}^{\pi} &=&
\left(
\begin{array}{cc}
  V_{\mathrm C}-V_{\mathrm T} & -2V_{\mathrm C}-V_{\mathrm T}    \\
  -2V_{\mathrm C}-V_{\mathrm T} & V_{\mathrm C}-V_{\mathrm T}
\end{array}
\right) \, ,
\label{eq:pi_pot_1-+} \\
V_{2^{+-}}^{\pi} &=&
\left(
\begin{array}{cc}
  V_{\mathrm C}-V_{\mathrm T} & -2V_{\mathrm C}-V_{\mathrm T}    \\
  -2V_{\mathrm C}-V_{\mathrm T} & V_{\mathrm C}-V_{\mathrm T}
\end{array}
\right) \, ,
\label{eq:pi_pot_2+-} \\
V_{2^{++}}^{\pi} &=&
\left(
\begin{array}{cccccc}
  0 & 0 & \sqrt{3} V_{\mathrm C} & -\sqrt{\frac{6}{5}} V_{\mathrm T} & 2\sqrt{\frac{3}{7}} V_{\mathrm T} & -6\sqrt{\frac{3}{35}} V_{\mathrm T} \\
  0 & -V_{\mathrm C}+V_{\mathrm T} & 0 & -3\sqrt{\frac{2}{5}} V_{\mathrm T} & \frac{3}{\sqrt{7}} V_{\mathrm T} & \frac{12}{\sqrt{35}} V_{\mathrm T} \\
  \sqrt{3} V_{\mathrm C} & 0 & 2 V_{\mathrm C} & \sqrt{\frac{2}{5}} V_{\mathrm T} & -\frac{2}{\sqrt{7}} V_{\mathrm T} & \frac{6}{\sqrt{35}} V_{\mathrm T} \\
  -\sqrt{\frac{6}{5}} V_{\mathrm T} & -3\sqrt{\frac{3}{5}} V_{\mathrm T} & \sqrt{\frac{2}{5}} V_{\mathrm T} & -V_{\mathrm C} & -\sqrt{\frac{14}{5}} V_{\mathrm T} & 0 \\
  2\sqrt{\frac{3}{7}} V_{\mathrm T} & \frac{3}{\sqrt{7}} V_{\mathrm T} & -\frac{2}{\sqrt{7}} V_{\mathrm T} & -\sqrt{\frac{14}{5}} V_{\mathrm T} & -V_{\mathrm C} - \frac{3}{7} V_{\mathrm T} & -\frac{12}{7\sqrt{5}} V_{\mathrm T} \\
  -6\sqrt{\frac{3}{35}} V_{\mathrm T} & \frac{12}{\sqrt{35}} V_{\mathrm T} & \frac{6}{\sqrt{35}} V_{\mathrm T} & 0 & -\frac{12}{7\sqrt{5}} V_{\mathrm T} & -V_{\mathrm C} + \frac{10}{7} V_{\mathrm T}
\end{array}
\right) \, ,
\label{eq:pi_pot_2++} \\
V_{2^{-+}}^{\pi} &=&
\left(
\begin{array}{cccc}
  V_{\mathrm C} + \frac{1}{5} V_{\mathrm T} & -\frac{3\sqrt{6}}{5} V_{\mathrm T} & -2V_{\mathrm C} + \frac{1}{5}V_{\mathrm T} & -\frac{3\sqrt{6}}{5} V_{\mathrm T}   \\
  -\frac{3\sqrt{6}}{5} V_{\mathrm T} & V_{\mathrm C}+\frac{4}{5}V_{\mathrm T}  & -\frac{3\sqrt{6}}{5} V_{\mathrm T} & -2V_{\mathrm C} + \frac{4}{5} V_{\mathrm T} \\
 -2V_{\mathrm C} + \frac{1}{5} V_{\mathrm T} & -\frac{3\sqrt{6}}{5} V_{\mathrm T}  & V_{\mathrm C} + \frac{1}{5} V_{\mathrm T} & -\frac{3\sqrt{6}}{5} V_{\mathrm T} \\
 -\frac{3\sqrt{6}}{5} V_{\mathrm T} & -2V_{\mathrm C} + \frac{4}{5} V_{\mathrm T} & -\frac{3\sqrt{6}}{5} V_{\mathrm T} & V_{\mathrm C} + \frac{4}{5} V_{\mathrm T}
\end{array}
\right) \, ,
\label{eq:pi_pot_2-+} \\
V_{2^{--}}^{\pi} &=&
\left(
\begin{array}{cccc}
 -V_{\mathrm C} - \frac{1}{5} V_{\mathrm T} & \frac{3\sqrt{6}}{5} V_{\mathrm T} & -\frac{3\sqrt{3}}{5} V_{\mathrm T} & \frac{6\sqrt{3}}{5} V_{\mathrm T}   \\
 \frac{3\sqrt{6}}{5} V_{\mathrm T} & -V_{\mathrm C}-\frac{4}{5}V_{\mathrm T}  & -\frac{3\sqrt{2}}{5} V_{\mathrm T} & \frac{6\sqrt{2}}{5} V_{\mathrm T} \\
 -\frac{3\sqrt{3}}{5} V_{\mathrm T} & -\frac{3\sqrt{2}}{5} V_{\mathrm T}  & -V_{\mathrm C} - \frac{7}{5} V_{\mathrm T} & -\frac{6}{5} V_{\mathrm T} \\
 \frac{6\sqrt{3}}{5} V_{\mathrm T} & \frac{6\sqrt{2}}{5} V_{\mathrm T} & -\frac{6}{5} V_{\mathrm T} & -V_{\mathrm C} + \frac{2}{5} V_{\mathrm T}
\end{array}
\right) \, .
\label{eq:pi_pot_2--}
\end{eqnarray}
}

\pagebreak

The $\rho$ and $\omega$ potentials are 

%
{\scriptsize
 \begin{align}
V^{v}_{0^{++}} &=
\left(
\begin{array}{ccc}
  V^{v \prime}_{\mathrm C} & 2\sqrt{3} V^{v}_{\mathrm C}  & \sqrt{6} V^{v}_{\mathrm T}   \\
  2\sqrt{3} V^{v}_{\mathrm C} & 4V^{v}_{\mathrm C} +V^{v \prime}_{\mathrm C} & -\sqrt{2} V^{v}_{\mathrm T}  \\
 \sqrt{6} V^{v}_{\mathrm T} & -\sqrt{2} V^{v}_{\mathrm T}  
&  -2V^{v}_{\mathrm C} -2V^{v}_{\mathrm T} +V^{v \prime}_{\mathrm C}
\end{array}
\right), \\
V^{v}_{0^{--}} &= 
\left(
-2V^{v}_{\mathrm C}+2V^{v}_{\mathrm T} +V^{v \prime}_{\mathrm C}
\right) , \\
V^{v}_{0^{-+}} &=
\left(
\begin{array}{cc}
  2V^{v}_{\mathrm C}-2V^{v}_{\mathrm T} +V^{v \prime}_{\mathrm C} & -4V^{v}_{\mathrm C}-2V^{v}_{\mathrm T}    \\
  -4V^{v}_{\mathrm C}-2V^{v}_{\mathrm T} & 2V^{v}_{\mathrm
   C}-2V^{v}_{\mathrm T} +V^{v \prime}_{\mathrm C}
\end{array}
\right) , \\
V^{v}_{1^{+-}} &=
\left(
\begin{array}{cccc}
  2V^{v}_{\mathrm C} +V^{v \prime}_{\mathrm C} & \sqrt{2}
   V^{v}_{\mathrm T} & -4V^{v}_{\mathrm C} & \sqrt{2} V^{v}_{\mathrm T}   \\
  \sqrt{2} V^{v}_{\mathrm T} & 2V^{v}_{\mathrm C}-V^{v}_{\mathrm T}+V^{v \prime}_{\mathrm C}  & \sqrt{2} V^{v}_{\mathrm T} & -4V^{v}_{\mathrm C} - V^{v}_{\mathrm T} \\
 -4V^{v}_{\mathrm C} & \sqrt{2} V^{v}_{\mathrm T}  & 
2V^{v}_{\mathrm C} +V^{v \prime}_{\mathrm C} & \sqrt{2} V^{v}_{\mathrm T} \\
 \sqrt{2} V^{v}_{\mathrm T} & -4V^{v}_{\mathrm C} - V^{v}_{\mathrm T} &
  \sqrt{2} V^{v}_{\mathrm T} & 2V^{v}_{\mathrm C} - V^{v}_{\mathrm T}
  +V^{v \prime}_{\mathrm C} 
\end{array}
\right) , \label{eq:V^{v}_{1^{+-}}} \\
V^{v}_{1^{++}} &=
\left(
\begin{array}{ccc}
  -2V^{v}_{\mathrm C}+V^{v \prime}_{\mathrm C} & -\sqrt{2}
   V^{v}_{\mathrm T} & -\sqrt{6}V^{v}_{\mathrm T} \\
  -\sqrt{2} V^{v}_{\mathrm T} & -2V^{v}_{\mathrm C} +V^{v}_{\mathrm T} &
  -\sqrt{3} V^{v}_{\mathrm T}  \\
 -\sqrt{6} V^{v}_{\mathrm T} & -\sqrt{3} V^{v}_{\mathrm T} &
  -2V^{v}_{\mathrm C}- V^{v}_{\mathrm T} +V^{v \prime}_{\mathrm C}
\end{array}
\right) , \\
V^{v}_{1^{--}} &=
\left(
\begin{array}{ccccc}
  V^{v \prime}_{\mathrm C} & 0 & 2\sqrt{3} V^{v}_{\mathrm C} &
   -2\sqrt{\frac{3}{5}}  V^{v}_{\mathrm T} & 3\sqrt{\frac{2}{5}}
   V^{v}_{\mathrm  T}   \\
  0 & -2V^{v}_{\mathrm C}-V^{v}_{\mathrm T} +V^{v \prime}_{\mathrm C} &
   0  & -3 \sqrt{\frac{3}{5}}
   V^{v}_{\mathrm T} &  3\sqrt{\frac{2}{5}} V^{v}_{\mathrm T} \\
 2\sqrt{3} V^{v}_{\mathrm C} & 0 & 4 V^{v}_{\mathrm C} +V^{v \prime}_{\mathrm C} &
  -\frac{4}{\sqrt{5}}V^{v }_{\mathrm C} & -\sqrt{\frac{6}{5}} V^{v}_{\mathrm T} \\
  -2\sqrt{\frac{3}{5}} V^{v}_{\mathrm T} & -3\sqrt{\frac{2}{5}}
   V^{v}_{\mathrm T} &  -\frac{4}{\sqrt{5}}V^{v}_{\mathrm C} &
   -2V^{v}_{\mathrm C} - \frac{7}{5} V^{v}_{\mathrm T} +V^{v
   \prime}_{\mathrm C} & \frac{\sqrt{6}}{5} V^{v}_{\mathrm T} \\
  3\sqrt{\frac{2}{5}} V^{v}_{\mathrm T} & -3\sqrt{\frac{2}{5}}
   V^{v}_{\mathrm T} & -\sqrt{\frac{6}{5}} V^{v}_{\mathrm T} &
   \frac{\sqrt{6}}{5} V^{v}_{\mathrm T} & -2V^{v}_{\mathrm C} -
   \frac{8}{5} V^{v}_{\mathrm T} +V^{v \prime}_{\mathrm C}
\end{array}
\right) , \\
V^{v}_{1^{-+}} &=
\left(
\begin{array}{cc}
  2V^{v}_{\mathrm C}+V^{v}_{\mathrm T} +V^{v \prime}_{\mathrm C}& -4V^{v}_{\mathrm C}+-V^{v}_{\mathrm T}    \\
  -4V^{v}_{\mathrm C}+V^{v}_{\mathrm T} & 2V^{v}_{\mathrm
   C}+V^{v}_{\mathrm T} + V^{v \prime}_{\mathrm C}
\end{array}
\right) , \\
V^{v}_{2^{+-}} &=
\left(
\begin{array}{cc}
  2V^{v}_{\mathrm C}+V^{v}_{\mathrm T} +V^{v \prime}_{\mathrm C}& -4V^{v}_{\mathrm C}+V^{v}_{\mathrm T}    \\
  -4V^{v}_{\mathrm C}+V^{v}_{\mathrm T} & 2V^{v}_{\mathrm
   C}+V^{v}_{\mathrm T} + V^{v \prime}_{\mathrm C}
\end{array}
\right) , \\
V^{v}_{2^{++}}  & = 
 \left( 
\begin{array}{cccccc}
  V^{v \prime}_{\mathrm C} & 0 & 2\sqrt{3} V^{v}_{\mathrm C} &
   \sqrt{\frac{6}{5}} V^{v}_{\mathrm T} & -2\sqrt{\frac{3}{7}}
   V^{v}_{\mathrm T} & 6\sqrt{\frac{3}{35}} V^{v}_{\mathrm T} \\
  0 & -2V^{v}_{\mathrm C}-V^{v}_{\mathrm T} +V^{v \prime}_{\mathrm C}& 0 & 3\sqrt{\frac{2}{5}}
   V^{v}_{\mathrm T} & -\frac{3}{\sqrt{7}} V^{v}_{\mathrm T} &  -\frac{12}{\sqrt{35}} V^{v}_{\mathrm T} \\
  2\sqrt{3} V^{v}_{\mathrm C} & 0 & 4 V^{v}_{\mathrm C} +V^{v
   \prime}_{\mathrm C} &
   -\sqrt{\frac{2}{5}} V^{v}_{\mathrm T} & \frac{2}{\sqrt{7}}
   V^{v}_{\mathrm T} &  -\frac{6}{\sqrt{35}} V^{v}_{\mathrm T} \\
  \sqrt{\frac{6}{5}} V^{v}_{\mathrm T} & -3\sqrt{\frac{3}{5}}
   V^{v}_{\mathrm T} &  -\sqrt{\frac{2}{5}} V^{v}_{\mathrm T} &
   -2V^{v}_{\mathrm C} +V^{v \prime}_{\mathrm C} & \sqrt{\frac{14}{5}} V^{v}_{\mathrm T} & 0 \\
  -2\sqrt{\frac{3}{7}} V^{v}_{\mathrm T} & -\frac{3}{\sqrt{7}}
   V^{v}_{\mathrm T} & \frac{2}{\sqrt{7}} V^{v}_{\mathrm T} &
   \sqrt{\frac{14}{5}} V^{v}_{\mathrm T} & -2V^{v}_{\mathrm C} +
   \frac{3}{7} V^{v}_{\mathrm T} +V^{v \prime}_{\mathrm C} & \frac{12}{7\sqrt{5}} V^{v}_{\mathrm T} \\
  6\sqrt{\frac{3}{35}} V^{v}_{\mathrm T} & -\frac{12}{\sqrt{35}}
   V^{v}_{\mathrm T}  & -\frac{6}{\sqrt{35}} V^{v}_{\mathrm T} & 0 &
   \frac{12}{7\sqrt{5}} V^{v}_{\mathrm T} & -2 V^{v}_{\mathrm C} -
   \frac{10}{7} V^{v}_{\mathrm T} +V^{v \prime}_{\mathrm C}
\end{array}
\right) , \\
V^{v}_{2^{-+}} &=
\left(
\begin{array}{cccc}
  2V^{v}_{\mathrm C} - \frac{1}{5} V^{v}_{\mathrm T} +V^{v
   \prime}_{\mathrm C} & \frac{3\sqrt{6}}{5} V^{v}_{\mathrm T} &
   -4V^{v}_{\mathrm C} - \frac{1}{5}V^{v}_{\mathrm T} & \frac{3\sqrt{6}}{5} V^{v}_{\mathrm T}   \\
  \frac{3\sqrt{6}}{5} V^{v}_{\mathrm T} & 2V^{v}_{\mathrm C}
   -\frac{4}{5}V^{v}_{\mathrm T} +V^{v \prime}_{\mathrm C}  & \frac{3\sqrt{6}}{5} V^{v}_{\mathrm T} & -4V^{v}_{\mathrm C} - \frac{4}{5} V^{v}_{\mathrm T} \\
 -4V^{v}_{\mathrm C} - \frac{1}{5} V^{v}_{\mathrm T} &
  \frac{3\sqrt{6}}{5} V^{v}_{\mathrm T}  & 2V^{v}_{\mathrm C} -
  \frac{1}{5} V^{v}_{\mathrm T} + V^{v \prime}_{\mathrm C} & \frac{3\sqrt{6}}{5} V^{v}_{\mathrm T} \\
 \frac{3\sqrt{6}}{5} V^{v}_{\mathrm T} & -4V^{v}_{\mathrm C} -
  \frac{4}{5} V^{v}_{\mathrm T} & \frac{3\sqrt{6}}{5} V^{v}_{\mathrm T}
  & 2V^{v}_{\mathrm C} - \frac{4}{5} V^{v}_{\mathrm T} + V^{v \prime}_{\mathrm C}
\end{array}
\right) , \\
V^{v}_{2^{--}}  &=
\left(
\begin{array}{cccc}
 -2V^{v}_{\mathrm C} + \frac{1}{5} V^{v}_{\mathrm T} + V^{v
  \prime}_{\mathrm C}  & -\frac{3\sqrt{6}}{5} V^{v}_{\mathrm T} &
  \frac{3\sqrt{3}}{5} V^{v}_{\mathrm T} & -\frac{6\sqrt{3}}{5} V^{v}_{\mathrm T}   \\
 -\frac{3\sqrt{6}}{5} V^{v}_{\mathrm T} & -2V^{v}_{\mathrm
  C}+\frac{4}{5}V^{v}_{\mathrm T} + V^{v \prime}_{\mathrm C} & \frac{3\sqrt{2}}{5} V^{v}_{\mathrm T} & -\frac{6\sqrt{2}}{5} V^{v}_{\mathrm T} \\
 \frac{3\sqrt{3}}{5} V^{v}_{\mathrm T} & \frac{3\sqrt{2}}{5}
  V^{v}_{\mathrm T}  &  -2V^{v}_{\mathrm C} + \frac{7}{5} V^{v}_{\mathrm
  T} + V^{v \prime}_{\mathrm C} & \frac{6}{5} V^{v}_{\mathrm T} \\
 -\frac{6\sqrt{3}}{5} V^{v}_{\mathrm T} & -\frac{6\sqrt{2}}{5}
  V^{v}_{\mathrm T} & \frac{6}{5} V^{v}_{\mathrm T} & -2V^{v}_{\mathrm
  C} - \frac{2}{5} V^{v}_{\mathrm T} + V^{v \prime}_{\mathrm C}
\end{array}
\right) .
\end{align}} 
where the central and tensor potentials  are defined as,
\begin{eqnarray}
V^{\pi}_{\mathrm C} &=& \left( \sqrt{2}\frac{g}{f_{\pi}}\right)^2
\frac{1}{3}C(r;m_{\pi})  \vec{\tau}_{1} \!\cdot\! \vec{\tau}_{2} \, , \\
V^{\pi}_{\mathrm T} &=& \left( \sqrt{2}\frac{g}{f_{\pi}}\right)^2
\frac{1}{3}T(r;m_{\pi})  \vec{\tau}_{1} \!\cdot\! \vec{\tau}_{2} \, , \\
V^{\rho}_{\mathrm C} &=& -\left( 2\lambda g_V \right)^2 \frac{1}{3}
C(r;m_{\rho}) \vec{\tau}_{1} \!\cdot\! \vec{\tau}_{2} \, , \\
V^{\omega}_{\mathrm C} &=&
 \left( 2\lambda g_V \right)^2 \frac{1}{3}C(r;m_{\omega}) \, ,  \\
V^{\rho}_{\mathrm T} &=& -\left( 2\lambda g_V \right)^2 \frac{1}{3}
T(r;m_{\rho}) \vec{\tau}_{1} \!\cdot\! \vec{\tau}_{2} \, , \\
V^{\omega}_{\mathrm T} &=&
 \left( 2\lambda g_V \right)^2 \frac{1}{3}T(r;m_{\omega}) \, ,  \\
 V^{\rho \prime}_{\mathrm C} &=& \left( \frac{\beta g_V}{2m_{\rho}} \right)^2
\frac{1}{3} C(r;m_{\rho})\vec{\tau}_{1} \!\cdot\! \vec{\tau}_{2} \, , \\
V^{\omega \prime}_{\mathrm C} &=&
- \left( \frac{\beta g_V}{2m_{\omega}} \right)^2
\frac{1}{3} C(r;m_{\omega}) \, . 
\end{eqnarray}

\section{Diagonalization of OPEP} \label{sec:OPEP}

We consider the diagonalization of OPEP (\ref{eq:pi_pot_0++})-(\ref{eq:pi_pot_2--}) by adopting a stationary approximation for $\mathrm{B}^{(\ast)}$ and $\bar{\mathrm{B}}^{(\ast)}$ mesons.
We regard the $\mathrm{B}^{(\ast)}$ and $\bar{\mathrm{B}}^{(\ast)}$ mesons as sources of isospin, and fix the positions of $\mathrm{B}^{(\ast)}$ and $\bar{\mathrm{B}}^{(\ast)}$ mesons by neglecting the kinetic term.
Then  the  potentials with channel-couplings as matrices  in which have
off-diagonal components,
turn to be diagonal matrices $\tilde{V}_{J^{PC}}^{\pi}$ as followings,
\begin{eqnarray}
\tilde{V}_{0^{++}}^{\pi} &=& \mbox{diag}\left( -3V_{\mathrm C}, -V_{\mathrm C}+4V_{\mathrm T}, -V_{\mathrm C}-2V_{\mathrm T} \right) \vec{\tau}_{1} \!\cdot\! \vec{\tau}_{2}, \\
\tilde{V}_{0^{--}}^{\pi} &=& \mbox{diag}\left(-V_{\mathrm C}-2V_{\mathrm T} \right) \vec{\tau}_{1} \!\cdot\! \vec{\tau}_{2}, \\
\tilde{V}_{0^{-+}}^{\pi} &=& \mbox{diag}\left( -3V_{\mathrm C}, -V_{\mathrm C}+4V_{\mathrm T} \right) \vec{\tau}_{1} \!\cdot\! \vec{\tau}_{2}, \\
\tilde{V}_{1^{+-}}^{\pi} &=& \mbox{diag}\left( 3V_{\mathrm C}, 3V_{\mathrm C}, -V_{\mathrm C}+4V_{\mathrm T}, -V_{\mathrm C}-2V_{\mathrm T} \right) \vec{\tau}_{1} \!\cdot\! \vec{\tau}_{2}, \\
\tilde{V}_{1^{++}}^{\pi} &=& \mbox{diag}\left( -V_{\mathrm C}+4V_{\mathrm T}, -V_{\mathrm C}-2V_{\mathrm T}, -V_{\mathrm C}-2V_{\mathrm T} \right) \vec{\tau}_{1} \!\cdot\! \vec{\tau}_{2}, \\
\tilde{V}_{1^{--}}^{\pi} &=& \mbox{diag}\left( 3V_{\mathrm C}, -V_{\mathrm C}+4V_{\mathrm T}, -V_{\mathrm C}+4V_{\mathrm T}, -V_{\mathrm C}-2V_{\mathrm T}, -V_{\mathrm C}-2V_{\mathrm T} \right) \vec{\tau}_{1} \!\cdot\! \vec{\tau}_{2}, \\
\tilde{V}_{1^{-+}}^{\pi} &=& \mbox{diag}\left( 3V_{\mathrm C}, -V_{\mathrm C}-2V_{\mathrm T} \right) \vec{\tau}_{1} \!\cdot\! \vec{\tau}_{2}, \\
\tilde{V}_{2^{+-}}^{\pi} &=& \mbox{diag}\left( 3V_{\mathrm C}, -V_{\mathrm C}-2V_{\mathrm T} \right) \vec{\tau}_{1} \!\cdot\! \vec{\tau}_{2}, \\
\tilde{V}_{2^{++}}^{\pi} &=& \mbox{diag}\left( 3V_{\mathrm C}, -V_{\mathrm C}-2V_{\mathrm T}, -V_{\mathrm C}-2V_{\mathrm T}, -V_{\mathrm C}-2V_{\mathrm T}, -V_{\mathrm C}+4V_{\mathrm T}, -V_{\mathrm C}+4V_{\mathrm T} \right) \vec{\tau}_{1} \!\cdot\! \vec{\tau}_{2}, \\
\tilde{V}_{2^{-+}}^{\pi} &=& \mbox{diag}\left( 3V_{\mathrm C}, 3V_{\mathrm C}, -V_{\mathrm C}-2V_{\mathrm T}, -V_{\mathrm C}+4V_{\mathrm T} \right) \vec{\tau}_{1} \!\cdot\! \vec{\tau}_{2}, \\
\tilde{V}_{2^{--}}^{\pi} &=& \mbox{diag}\left( -V_{\mathrm C}-2V_{\mathrm T}, -V_{\mathrm C}-2V_{\mathrm T}, -V_{\mathrm C}-2V_{\mathrm T}, -V_{\mathrm C}+4V_{\mathrm T} \right) \vec{\tau}_{1} \!\cdot\! \vec{\tau}_{2},
\end{eqnarray}
where the central and tensor potentials are defined as,
$V_{\mathrm C} = \left( \sqrt{2}\frac{g}{f_{\pi}} \right)^{2} \frac{1}{3} C(r;m_{\pi})$ and
$V_{\mathrm T} = \left( \sqrt{2}\frac{g}{f_{\pi}} \right)^{2} \frac{1}{3} T(r;m_{\pi})$,
with $V_{\mathrm C} > 0$ and $V_{\mathrm T}>0$  and $V_{\mathrm C} < V_{\mathrm T}$.
For $I=1$ ($\vec{\tau}_{1} \!\cdot\! \vec{\tau}_{2}=1$), we see that the strongest attractive potential, $-(V_{\mathrm C}+2V_{\mathrm T})$, is contained in the $J^{PC}=0^{++}$, $0^{--}$, $1^{+-}$, $1^{++}$, $1^{--}$, $1^{-+}$, $2^{+-}$, $2^{++}$, $2^{-+}$ and $2^{--}$ states.
In another quantum number, the  $0^{-+}$ state in $I=1$ has only weakly attractive potential, $-3V_{\mathrm C}$.
Therefore we expect in $I=1$ that the $0^{++}$, $0^{--}$, $1^{+-}$, $1^{++}$, $1^{--}$, $1^{--}$, $1^{-+}$, $2^{+-}$, $2^{++}$, $2^{-+}$ and $2^{--}$ states may be bound and/or resonant states, while the $0^{-+}$ state hardly forms a structure.
For $I=0$ ($\vec{\tau}_{1} \!\cdot\! \vec{\tau}_{2}=-3$), the strongest attractive potential, $-3 \cdot 3V_{\mathrm C}$, is contained in $1^{-+}$ and $2^{+-}$ states.
The potential in the $0^{--}$ state is repulsive.
Therefore there may be a bound and /or resonant states in $1^{-+}$ and $2^{+-}$, and no structure in $0^{--}$ in $I=0$.

Although the static approximation may  be a crude approximation, 
this is a useful method at qualitative level.
For example, let us study two nucleon ($\mathrm{NN}$) systems,
we analyze the deuteron ($I=0$, $J^{P} = 1^{+}$) in which the $\mathrm{NN}$ potential is given by $2 \times 2$ matrix with $^{3}S_{1}$ and $^{3}D_{1}$ states.
\footnote{There is also analysis for $\mathrm{N}\bar{\mathrm{N}}$ systems in Ref.~\cite{Dover:1978br}}
The OPEP for deuteron is given as
\begin{eqnarray}
V^{\pi}_{\mathrm{NN}} =
\left(
\begin{array}{cc}
 V^{\mathrm{NN}}_{\mathrm C} & 2\sqrt{2}V^{\mathrm{NN}}_{\mathrm T}  \\
 2\sqrt{2}V^{\mathrm{NN}}_{\mathrm T} & V^{\mathrm{NN}}_{\mathrm C}-2V^{\mathrm{NN}}_{\mathrm T}
\end{array}
\right)  \vec{\tau}_{1} \!\cdot\! \vec{\tau}_{2},
\label{eq:NN_deuteron}
\end{eqnarray}
with 
$V^{\mathrm{NN}}_{\mathrm C} = \left( \frac{g_{\pi \mathrm{NN}}}{2m_{N}} \right)^{2} \frac{1}{3} C(r;m_{\pi})$ and
$V^{\mathrm{NN}}_{\mathrm T} = \left( \frac{g_{\pi \mathrm{NN}}}{2m_{N}} \right)^{2} \frac{1}{3} T(r;m_{\pi})$,
for a $\pi \mathrm{NN}$ vertex constant $g_{\pi \mathrm{NN}}$ and a nucleon mass $m_{\mathrm{N}}$.
We diagonalize Eq.~(\ref{eq:NN_deuteron}) and obtain the eigenvalues in the diagonal potential,
\begin{eqnarray}
\tilde{V}^{\pi}_{\mathrm{NN}} = \mbox{diag} \left( V^{\mathrm{NN}}_{\mathrm C}-4V^{\mathrm{NN}}_{\mathrm T}, V^{\mathrm{NN}}_{\mathrm C}+2V^{\mathrm{NN}}_{\mathrm T} \right) \vec{\tau}_{1} \!\cdot\! \vec{\tau}_{2}.
\end{eqnarray}
Because $V^{\mathrm{NN}}_{\mathrm T}>V^{\mathrm{NN}}_{\mathrm C}$, the first eigenvalue gives a repulsion in $I=0$, while the second eigenvalue gives an attraction in $I=0$.
The eigenvector of the $^{3}S_{1}$ and $^{3}D_{1}$ components for the second eigenvalue is $(\sqrt{2/3},\sqrt{1/3})$.
This means that the $D$-wave probability is about 33 percents.
In reality, we have to introduce a kinetic term which disfavors the higher angular momentum ($D$-wave), and hence the $D$-wave probability in deuteron becomes a few ($\sim 5$) percents.
The stationary approximation will be applied to $\mathrm{B}^{(\ast)}\bar{\mathrm{B}}^{(\ast)}$ systems with better accuracy, because the $\mathrm{B}$ meson mass is 5.6 times larger than the nucleon mass.
We note that the OPEP as a long range force are better approximation for larger angular momentum, because the sizes of systems are extended.
However, the static approximation may become worse for larger angular momentum.
Therefore, it is necessary for quantitative analysis to study numerically the solutions of the Schr\"odinger equations with the kinetic terms and the potentials with channel-coupling as discussed in the text.

\end{document}